\documentclass{emulateapj}
\usepackage{apjfonts}[9pt]
\usepackage{graphicx}

\begin{document}

\title{Deep $u^*$- and $g$-band Imaging of
 the \textit{Spitzer Space Telescope} First Look Survey Field:
 Observations and Source Catalogs}

\author{Hyunjin Shim\altaffilmark{1,5}, Myungshin Im\altaffilmark{1,6},
 Soojong Pak\altaffilmark{2,4},
 Philip Choi\altaffilmark{3}, Dario Fadda\altaffilmark{3},
 George Helou\altaffilmark{3},
 \& Lisa Storrie-Lombardi\altaffilmark{3}}

\altaffiltext{1}{Dept. of Physics \& Astronomy, Seoul National University,
 Seoul, Korea}
\altaffiltext{2}{Korea Astronomy \& Space Science Institute, Daejeon, Korea}
\altaffiltext{3}{Spitzer Science Center, California Institute of Technology,
 Pasadena, CA 91125}
\altaffiltext{4}{Dept. of Astronomy and Space Science, Kyung Hee University,
Yongin-shi, Kyunggi-do 449-701, Korea}
\altaffiltext{5}{hjshim@astro.snu.ac.kr}
\altaffiltext{6}{mim@astro.snu.ac.kr}

  \begin{abstract}
  We present deep $u^*$-, and $g$-band images taken with the
 MegaCam on the 3.6 m Canada-France-Hawaii Telescope (CFHT) to
 support the extragalactic component of the \textit{Spitzer}
 First Look Survey (hereafter, FLS).
  In this paper we outline the observations, present source
 catalogs and characterize the completeness, reliability, 
 astrometric accuracy and number counts of this dataset. 
  In the central 1 deg$^2$ region of the FLS, we reach depths
 of $g\sim$26.5 mag, and $u^{*}\sim$26.2 mag  
 (AB magnitude, 5$\sigma$ detection over a 3$\arcsec$ aperture) 
 with $\sim$4 hours of exposure time for each filter.
  For the entire FLS region ($\sim$5 deg$^2$ coverage),
 we obtained $u^*$-band images to the shallower depth of 
 $u^{*}=$25.0--25.4 mag (5$\sigma$, 3$\arcsec$ aperture).
  The average seeing of the observations is 0$\farcs$85 for the 
 central field, and $\sim$1$\farcs$00 for the other fields. 
  Astrometric calibration of the fields yields an absolute 
 astrometric accuracy of 0$\farcs$15 when matched with 
 the SDSS point sources between 18$<g<$22.
  Source catalogs have been created using \textit{SExtractor}.
  The catalogs are 50\% complete and greater than 99.3\% reliable
 down to $g\simeq26.5$ mag and $u^{*}\simeq26.2$ mag for the 
 central 1 deg$^2$ field. In the shallower $u^*$-band images, the
 catalogs are 50\% complete and 98.2\% reliable down to 24.8--25.4
 mag.
  These images and source catalogs will serve as a useful resource 
 for studying the galaxy evolution using the FLS data.
  \end{abstract}


\section{INTRODUCTION}

  Since its launch in August 2003, the \textit{Spitzer} space 
 telescope has opened new IR observing windows to the universe
 and probed to unprecedented depths.
  The first scientific observation undertaken with the
 \textit{Spitzer} after its in-orbit-checkout period was the
 \textit{Spitzer} First Look Survey. 
  The survey  provided the first look of the mid-infrared (MIR)
 sky with deeper sensitivities than previous systematic large
 surveys\footnote{See the First Look Survey 
 website, http://ssc.spitzer.caltech.edu/fls}, allowing future 
 users to gauge \textit{Spitzer} sensitivities.
  For the extragalactic component of the First Look Survey (FLS),
 the $\sim$4.3 deg$^2$ field centered at RA=17$^h$18$^m$00$^s$,
 DEC=+59$\degr$30$\arcmin$00$\arcsec$ was observed with the
 InfraRed Array Camera (IRAC, 3.6, 4.5, 5.8, 8.0$\mu$m;
 Fazio et al. 2004) and the Multiband Imaging Photometer for Spitzer
 (MIPS, 24, 70, 160$\mu$m; Rieke et al. 2004).
  The main survey field has flux limits of
 20, 25, 100, and 100$\mu$Jy at wavelengths of IRAC 3.6, 4.5, 5.8, 
 and 8.0$\mu$m respectively.
  The deeper verification field, covering the central 900 arcmin$^2$
 field,  has sensitivities of 10, 10, 30, and 30$\mu$Jy at IRAC
 wavelengths. (Lacy et al. 2005)
  Also, the FLS verification strip field has 3$\sigma$ flux limits of 
 90$\mu$Jy at MIPS 24$\mu$m (Yan et al. 2004a), and 
 9, and 60 mJy at 70, 160$\mu$m (Frayer et al. 2005).  
  The reduced images and source catalogs have been released 
 (Lacy et al. 2005; Frayer et al. 2005).

  In order to support the FLS, various ground-based ancillary datasets 
 have also been obtained. 
  A deep $R$-band survey to $R_{AB}$(5$\sigma$)=25.5 mag
 (KPNO 4 m, Fadda et al. 2004) and a radio survey at 1.4GHz to
 90$\mu$Jy (VLA, Condon et al. 2003) have been performed to identify
 sources at different wavelengths.
  Imaging surveys with the Palomar Large Format Camera (LFC), 
 $g^{\prime}$, $r^{\prime}$, $i^{\prime}$ and $z^{\prime}$ band data 
 covering 1--4 deg$^2$ have also been taken since 2001.
  The Sloan Digital Sky Survey also covers the FLS field.
 Moreover, spectroscopic redshifts have been obtained for many
 sources using the DEep Imaging Multi-Object Spectrograph (DEIMOS) 
 at the Keck observatory as well as Hydra instrument on 
 the Wisconsin Indiana Yale NOAO (WIYN) observatory.
  Utilizing all the \textit{Spitzer} and the ground-based ancillary
 datasets, many science results have come out already from the FLS.
 The FLS studies range from the infrared source counts to the
 properties of obscured galaxies such as submm galaxies, red Active
 Galactic Nuclei(AGNs), and Extremely Red Objects
 (e.g. Yan et al. 2004b; Fang et al. 2004; Marleau et al. 2004;
 Frayer et al. 2004; Appleton et al. 2004; Lacy et al. 2004;
 Choi et al. 2005).

  In order to broaden the scientific scope of the FLS, we have
 obtained new ground-based ancillary datasets, deep $u^*$-band
 and $g$-band imaging data that will be presented in this paper. 
  The exploration of the cosmic star formation history at $z\lesssim3$ 
 is one of the key scientific issues that the FLS is designed
 to address.
  While studying the infrared emission is an extremely valuable way   
 to understand the obscured star formation history of the universe,
 a complete census of the cosmic star formation history requires 
 the measurement of unobscured star formation as well. 
  Such a measurement can be done effectively using the rest-frame
 ultraviolet (UV) continuum, which represents either instantaneous
 star formation activity of massive stars (Far UV; FUV), or 
 traces the star formation from intermediate age stars (Near UV; NUV)
 (Kennicutt 1998). The FUV and NUV emission of galaxies at
 $z\lesssim1$ redshifts to the $u$-band; therefore, the addition
 of the $u$-band enables us to describe the star formation history
 at $z\lesssim1$ to the full extent. 
  Also, using the deep $u$-band and $g$-band data, it is possible
 to select $u$-band dropout galaxies --  star forming galaxies at
 $z\sim3$. 
  Especially of interest are bright, massive Lyman Break Galaxies
 which are rare, but important to constrain galaxy evolution models 
 since they can tell us how massive galaxies are assembled in the early 
 universe. The wide area coverage of the FLS allows us to select
 such rare, bright LBGs. However, for the selection of LBGs, it is
 necessary to obtain $u$-band data and $g$-band data that matches
 the depth of the \textit {Spitzer} data.
  In addition, the $u$-band and $g$-band datasets can be used to test
 extinction correction methods, such as the estimation of dust extinction
 based on the measurement of UV slope (Calzetti et al. 1994;
 Meurer et al. 1999; Adelberger \& Steidel, 2000).
  When used together with other ground-based ancillary data, our $u$-
 and $g$-band data will be very helpful for improving the accuracy of
 photometric redshifts. 
 
  With these scientific applications in mind, we present the $u^*$-
 and $g$-band optical observations made with the MegaCam at
 Canada France Hawaii Telescope (CFHT). 
  The dataset is composed of  
  i)  deep $u^*$-, $g$-band data for the central 1 deg$^2$
 of the FLS, and
  ii) $u^*$-band data for the whole FLS.
  In \S2, we describe how our observations were made. The reduction and 
 calibration procedures are in \S3. Production of source catalogs 
 is described in \S4. Finally in \S5, we show the properties of the
 final images and the extracted catalogs. Throughout this paper, we use
 AB magnitudes (Oke 1974).

\section{OBSERVATIONS}

\begin{deluxetable}{ccc}
\tabletypesize{\small}

\tablewidth{0pt}
\tablecaption{Characteristics of the filters used in MegaCam observations\label{tab:filter}}
\tablehead{
\colhead{Filter} & \colhead{$u^{*}$} & \colhead{$g$}
}
\startdata
  central wavelength ($\mbox{\AA}$) & 3740 & 4870 \\
  wavelength range & 3370 -- 4110 & 4140 -- 5590 \\
  mean transmission (\%) & 69.7 & 84.6 \\
\enddata
\end{deluxetable}

\begin{figure}
     \epsscale{1.0}
     \plotone{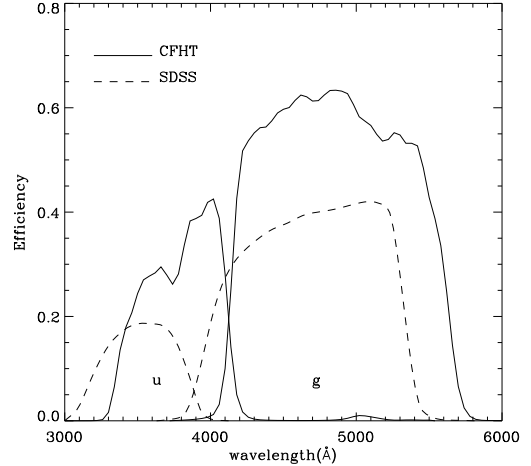}
     \caption{\label{fig:filter} The response curves of CFHT $u^*$, $g$
     filters multiplied by the telescope (mirror, optics) and CCD
     efficiency (\textit{solid}). The CFHT data points are taken
     from a CFHT-related homepage
     (http://astrowww.phys.uvic.ca/grads/gwyn/cfhtls).
     The SDSS $u$-, $g$-band transmission curves are
     given as \textit{dashed} lines (http://www.sdss.org).
     The CFHT filters are slightly redder than their SDSS counterparts.}
     \end{figure}

  The observations were made using the MegaCam on the 3.6 m Canada
 France Hawaii Telescope (CFHT). MegaCam consists of 36 2048$\times$4612
 pixel CCDs, covering 0.96 deg$\times$0.94 deg field of view with
 a resolution of 0.187$\arcsec$ per pixel (See the detailed description
 in Boulade et al. 2003).
  The available broad-band filters are similar to SDSS \textit{ugriz}
 filters, but not exactly the same especially in the case of
 $u^*$-band. We used $u^*$ and $g$ filters for our observations.
 The filter characteristics are summarized in Table {\ref{tab:filter}},
 and their response curves are compared to the SDSS $u$- and $g$-band
 filters in Figure {\ref{fig:filter}}. Note that these response curves
 include the quantum efficiency of the CCDs.
  The response curves of the two $g$-bands are nearly identical in their
 shapes, but the $u^*$-band response curve shows a different overall
 shape compared to the SDSS $u$-band, in such a sense that $u^*$ filter
 is redder than SDSS $u$.  
  The difference leads to as much as a 0.6 mag difference between the
 CFHT \& SDSS $u$-magnitudes. 
  This will be investigated further in the photometry section (\S5.2).

     \begin{figure*}
     \epsscale{.9}
     \plotone{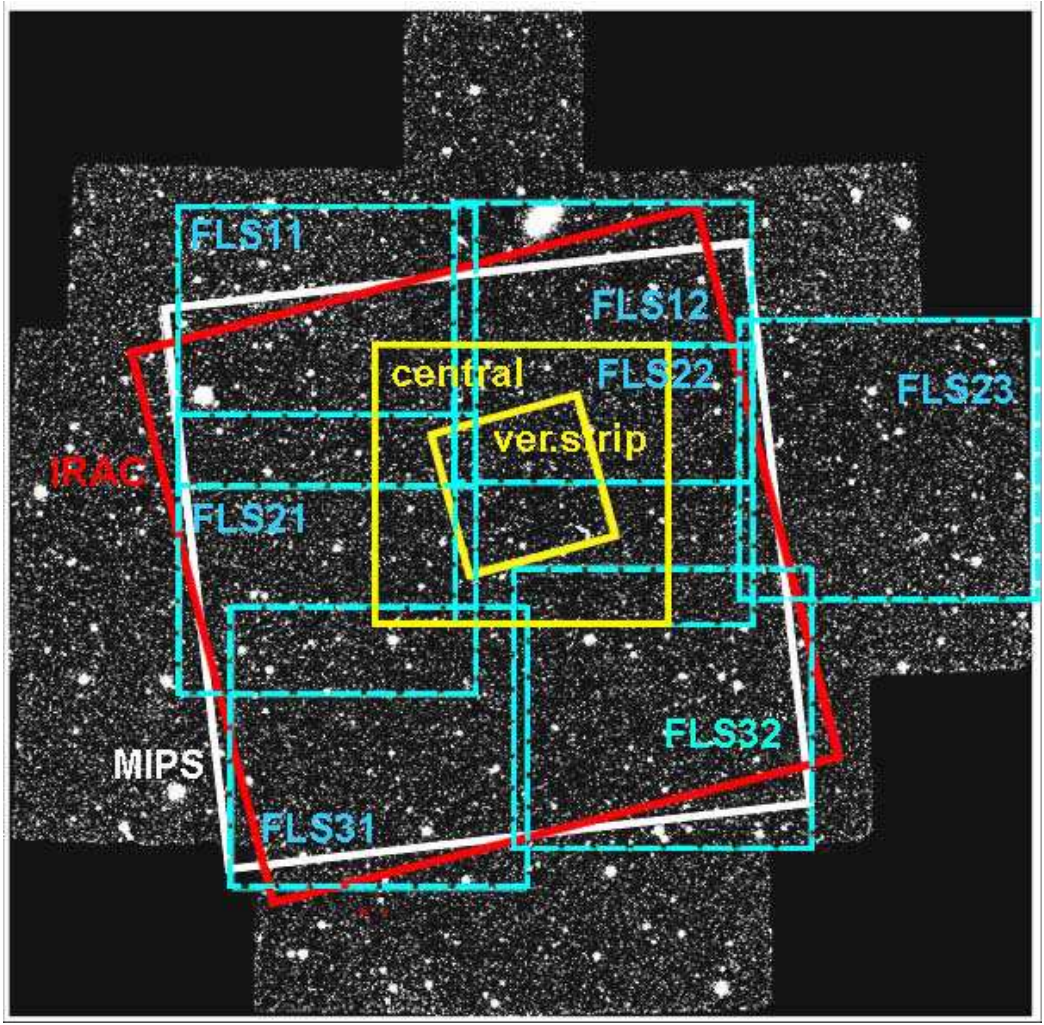}
     \caption{\label{fig:area} Area coverages of various datasets
     over the FLS. The small yellow square in the center represents
     the \textit{Spitzer} FLS verification strip, and the big yellow square
     shows our CFHT observations in the central 1 deg$^2$.
     The red/white squares correspond to the whole FLS area (IRAC/MIPS
     respectively) which covers $\sim$4.3 deg$^2$. The cyan squares indicate
     the area covered by our CFHT $u^*$-band observations in 2004.
     All these marks are overlayed on the KPNO $R$-band images.}
     \end{figure*}

  Our images on the FLS field were taken with the queued observation
 mode in two separate runs (03B and 04A semesters). 
  The first observing run occurred in August 23--29 ($u^*$-band) and
 September 19--22 ($g$-band), 2003. During this 03B run, we observed
 the central 1 deg$^2$ of the FLS field, which covers the entire FLS
 verification strip.
  Images are taken with medium dither steps of about 30$\arcsec$, at
 12 ($u^*$) or 20 ($g$) different dither positions.
  The exposure time at each position was 1160 seconds for $u^*$-band, 
 680 seconds for $g$-band.
  Therefore, the total integration time per pixel amounts to $\sim$3.8
 hours for both bands. 
  The yellow-colored box in Figure {\ref{fig:area}}, shows the location
 of this deep $u^*$- and $g$-band data superimposed on the $R$-band
 images (Fadda et al. 2004).

  In the second observing run during 04A period, we obtained $u^*$-band
 images covering the whole FLS field ($\sim$5 deg$^2$) at a shallower
 depth. Note that we did not take $g$-band data because the entire FLS
 is covered to a shallower depth by the existing $g$-band data taken
 with the Palomar 5 m telescope. 
  The CFHT observation was at first designed to take 10 dithered images
 at 7 different locations of the FLS area, which are named as FLS11--FLS32
 (Table {\ref{tab:obslog}}). The cyan-colored boxes marked with the
 dashed line in Figure {\ref{fig:area}} indicate FLS11--FLS32 fields.
  The dates when the observations made span quite a wide range from
 April to July 2004. Some of the fields have all dithered images 
 observed on a photometric condition (FLS21, FLS22 \& FLS31), but other
 fields contain the images that are non-photometric.
  Due to the bad weather, we lost some of the queued observing
 time and several observed images were not validated. 
  Each dithered image was taken with the exposure time of 680 seconds.
 Although the original plan was to achieve 1.9 hours of total integration
 at each location, there is a field-to-field variation on the image depth
 between $\sim$1 hour and $\sim$2 hours.
  Table {\ref{tab:obslog}} gives the summary of the 03B and 04A
 observations. The seeing condition of the whole observation sessions
 ranged from 0$\farcs$85 to 1$\farcs$08. 

\begin{deluxetable*}{ccccccccc}
\tabletypesize{\scriptsize}
\tablecaption{\label{tab:obslog} Observation summary}
\tablewidth{0pt}
\tablehead{
\colhead{Field ID} & \colhead{$\alpha$ (J2000)} & \colhead{$\delta$ (J2000)} 
& \colhead{Observation Dates}
& \colhead{Filter} & \colhead{Total Exposure Time (sec)} & \colhead{Depth (AB mag)\tablenotemark{a}}
& \colhead{Seeing ($\arcsec$)\tablenotemark{b}} & \colhead{Zeropoint\tablenotemark{c}}
}
\startdata
 central & 17:17:01 & +59:45:08 & 2003/08/25, 2003/08/29 & $u^*$ & 1160 $\times$ 12 & 26.41 & 0.85 & 32.904 \\
 central & 17:17:01 & +59:45:08 & 2003/09/19, 2003/09/21, 2003/09/22 & $g$ & 680 $\times$ 20 & 26.72 & 0.85 & 33.430 \\
 \tableline
 FLS11 & 17:22:09 & +60:14:47 & 2004/04/29, 2004/07/12 & $u^*$ & 680 $\times$ 10 & 25.83 & 0.98 & 32.226 \\
 FLS12 & 17:14:44 & +60:14:50 & 2004/06/18, 2004/07/17 & $u^*$ & 680 $\times$ 9  & 25.85 & 1.06 & 32.186 \\
 FLS21 & 17:22:14 & +59:30:00 & 2004/07/10, 2004/07/21 & $u^*$ & 680 $\times$ 9  & 25.91 & 0.98 & 32.236 \\
 FLS22 & 17:14:47 & +59:44:45 & 2004/07/10 & $u^*$ & 680 $\times$ 5  & 25.57 & 1.08 & 32.213 \\
 FLS23 & 17:07:09 & +59:48:26 & 2004/07/11 & $u^*$ & 680 $\times$ 5  & 25.43 & 0.89 & 32.221 \\
 FLS31 & 17:20:46 & +58:49:09 & 2004/07/11, 2004/07/22 & $u^*$ & 680 $\times$ 8  & 25.68 & 0.90 & 32.188 \\
 FLS32 & 17:13:32 & +58:57:46 & 2004/07/12 & $u^*$ & 680 $\times$ 7  & 25.67 & 0.95 & 32.139 \\
\enddata
\tablenotetext{a}{The depth of a field is calculated as 5$\sigma$ flux over a 3$\arcsec$ aperture.}
\tablenotetext{b}{The seeing of the image was determined through a visual inspection on individual stars.}
\tablenotetext{c}{The zeropoints are calibrated from original Elixir photometric solution (See \S3.5). The values are given as zeropoints according to DN.}
\end{deluxetable*}

  Between August 2003 and September 2003, one of the 36 MegaCam
 CCDs was malfunctioning. As a result, $g$-band images lack [CCD03]
 part of the observed field.
  Hence, the effective survey area in coadded mosaic image is
 $\sim$0.94 deg$^2$ for the central field.

 \section{DATA REDUCTION}

  \begin{figure*}
  \epsscale{1.1}
  \plotone{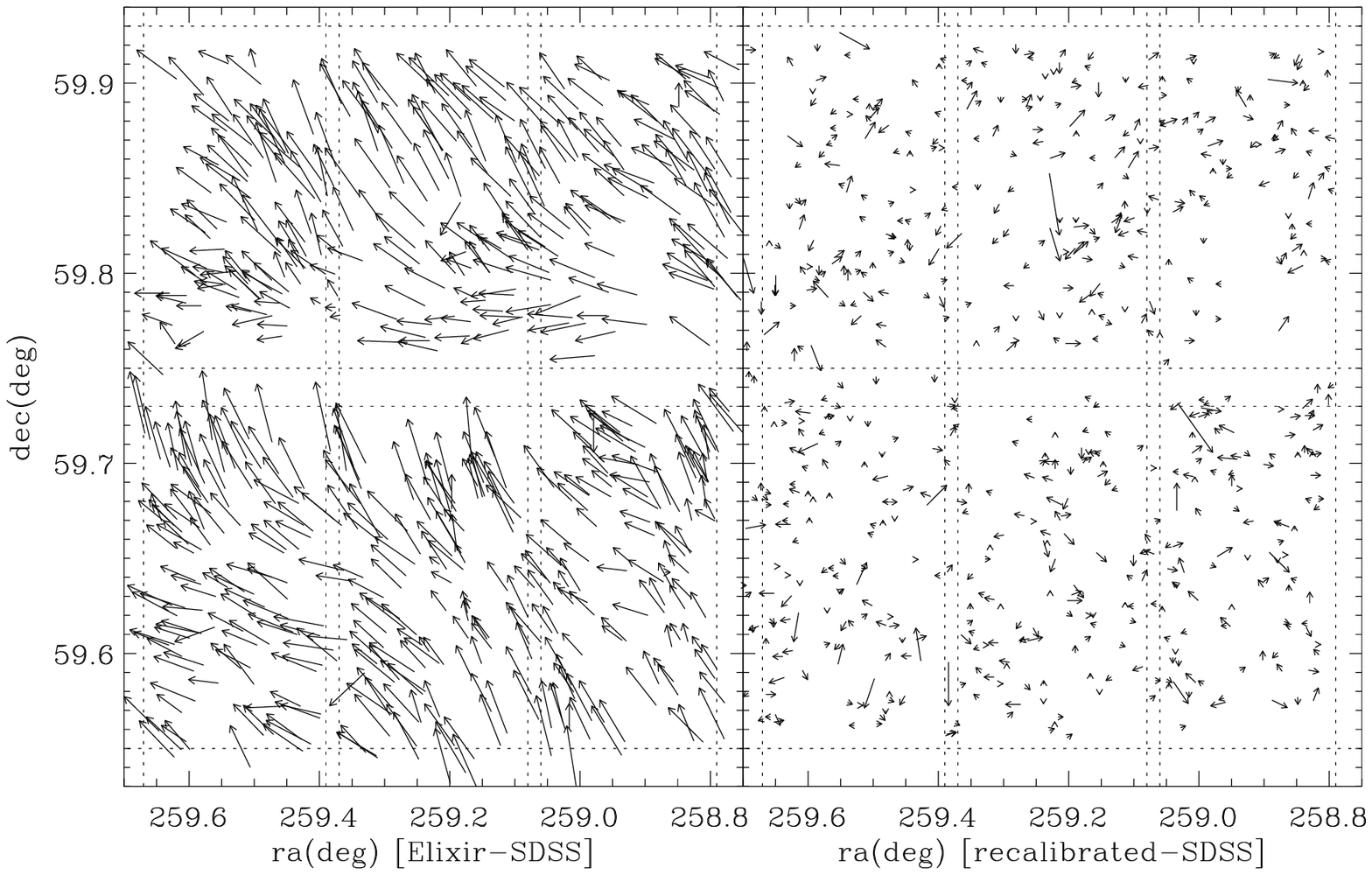}
     \caption{\label{fig:astrom_imp} The figure shows the improvement
 in astrometric solutions after the astrometric calibration with SDSS
 sources. The vectors indicate the difference between reference SDSS
 positions vs detected positions. The size of the vectors are
 exaggerated by a factor of 250.
    The head of the arrows are pointing to SDSS positions.
    After the calibration, the size of the vectors remarkably
 decreased (more than a factor of 5) and the distortion is reduced
 significantly.
    The figure represents 6 adjacent CCD chips.
    The borders of the CCD chips are shown as dashed lines.}
 \end{figure*}

  \subsection{Preprocessing with Elixir}  

  All MegaCam data obtained at CFHT are preprocessed through the
 Elixir pipeline\footnote{See Elixir website
 http://www.cfht.hawaii.edu/Instruments/Elixir}, the standard
 pipeline for the basic reduction of the MegaCam data, provided
 by the CFHT.
  Through the Elixir pipeline, the raw object frames are bias
 subtracted, flat fielded, and photometrically calibrated. In the
 Elixir photometric calibration, the nightly photometric zeropoint
 is estimated using the standard star data on that night. 
  Basic astrometric calibrations are done using $\sim$50--60  
 USNO (United States Naval Observatory catalog; Zacharias et al.
 2000) and HST GSC (Hubble Space Telescope Guide Star Catalog;
 Morrison et al. 2001) reference stars per chip. 
  In this calibration, the average pixel scale and the image
 rotation are derived and written as WCS keywords to the header
 of the raw image. 
  Since the Elixir astrometric solution uses only a first order
 fit,  the absolute astrometric accuracy of the Elixir-processed
 images is about 0$\farcs$5--1$\arcsec$ rms with respect to the
 reference stars above. 
  Final processed images are delivered to us as Multiple Extension
 Fits files 
 that can be manipulated using IRAF\footnote{IRAF is distributed
 by the National Optical Astronomy Observatories, which are operated
 by the Association of Universities for Research in Astronomy, Inc.,
 under cooperative agreement 
 with the National Science Foundation.} \texttt{mscred}.

  \subsection{Post-Elixir Processing before Mosaicking}  

  The delivered images are processed before mosaicking,  
 using various IRAF packages and tasks. 
  The post-Elixir image processing includes: (1) identification of 
 saturated pixels and bleed trails; (2) creation of new bad pixel masks;
 (3) removal of satellite trails by inspecting each image frame;
 (4) replacement of bad pixel values; and (5) removal of cosmic rays.

  As the first stage of the post-Elixir image processing, we
 identified saturated pixels and bleed trails using 
 the \texttt{ccdproc} task in IRAF. 
  The saturated pixels and bleed trails, which are the pixels
 showing nonlinear behavior, should be identified before stacking
 images because they can affect surrounding pixels during the
 projection of the image. 
  Those pixels with values above the saturation level, 64000 DN
 for $u^*$-band and 62000 DN for $g$-band, were identified as
 saturated pixels. 
  Neighboring pixels within a distance of 5 pixels along lines or
 columns of the selected saturation pixels are also identified as
 saturated. We found that the pixel saturation occurs typically at
 the central part of the stars brighter than $u^{*}\simeq$20 mag and
 $g\simeq$21 mag. 
  Bleed trails, which result from charge spillage from a CCD pixel
 above its capacity, tend to run down the columns. 
  We identify bleed trails as more than 20 connected pixels that
 are 5000 counts above the mean value along the whole image (i.e.,
 setting the \texttt{ccdproc} threshold parameter as ``mean+5000''). 
 After the identification of saturated pixels and bleed trails, 
 this information is stored as a saturation mask.
  
  Next, we used the IRAF task \texttt{imcombine} to make new bad
 pixel masks by combining the saturation masks with the original
 CFHT bad pixel masks. 
  The resultant masks have pixel values of 0 for good pixels,
 greater than 0 for bad pixels.
  In the bad pixel masks, we also indicate the location of
 satellite trails.
  To include satellite trails in bad pixel masks, we identified
 satellite trails through the visual inspection by flagging their
 start/end (x, y) image coordinates and widths on each CCD chip.
  The flagged rectangular line is then marked as bad pixels in
 the bad pixel masks.  

  With the newly constructed bad pixel masks, we fixed the bad
 pixels using the IRAF \texttt{fixpix} task. 
  The value of the pixel marked in the bad pixel masks was replaced
 by the linear interpolation value from adjacent pixels. 
  The next step is to identify the pixels affected by cosmic rays.
  We applied a robust algorithm based on a variation of Laplacian
 edge detection with the program, \texttt{la\_cosmic} (van Dokkum 2001)
 on each CCD frame. This algorithm identifies cosmic rays of arbitrary
 sizes and shapes, and our visual inspection of this procedure confirms
 effective removal of cosmic rays. 
  After the cosmic ray cleaning, we improved the astrometric solution,
 which is described in the next section.

  \subsection{Astrometric Calibration}

  The Elixir data pipeline performs a rough astrometric calibration
 on each image, and the existing header WCS keywords are updated
 with the improved astrometric solution. As mentioned in Elixir
 section (\S3.1),  Elixir calculates the astrometric solution for
 each CCD chip. An initial guess is determined for the image coordinate
 based on the initial header WCS keywords, and Elixir uses HST Guide
 Star Catalogs or USNO database to calculate an astrometric solution. 
  The solution derived by Elixir is a first order fit that gives the
 rms scatter of 0$\farcs$5$\sim$1$\arcsec$. Thus, the solution is not
 good enough for follow-up observations that require very accurate
 position information such as the multi-object spectroscopy.
  The accurate astrometry is also important when combining and
 mosaicking images, since misalignment of images can lead to a loss
 of signal in the combined image.

  To improve the astrometric solutions, we adopted SDSS sources
 as reference points, and derived the astrometric solution using
 \texttt{msctpeak} task in \texttt{mscfinder}, which is a subpackage
 of the IRAF mosaic reduction package, \texttt{mscred}.
  The SDSS sources brighter than $r\simeq$20 mag have an absolute
 astrometric accuracy of 0$\farcs$045 rms with respect to USNO catalogs
 (Zacharias et al. 2000) and 0$\farcs$075 rms against Tycho--2 catalogs
 (Hog et al. 2000) with an additional 0$\farcs$02$\sim$0$\farcs$03
 systematic error in both cases (Pier et al. 2003).
  We also considered using USNO and GSC catalogs; unfortunately, many
 of the USNO/GSC stars were saturated in our images, making it difficult
 to use them for astrometric calibration.
  
  There are other tasks such as \texttt{msczero} and \texttt{msccmatch} 
 for deriving the astrometric solution, but we settled on using 
 \texttt{msctpeak}.
  Solutions from tasks such as \texttt{msccmatch} -- which works only
 in \texttt{TAN} projection -- are reflected in the value of
 \texttt{CDi\_j}, \texttt{CRPIXi} header keywords. As a result, the 
 astrometric solutions from such tasks have limitations and they do not
 improve the astrometric accuracy significantly. 
  On the other hand, \texttt{msctpeak} creates a separate WCS database
 including a higher order fit solution from \texttt{TNX} sky projection
 geometry, and they are written as additional keywords such as
 \texttt{WATi\_00j} in the FITS header using \texttt{mscsetwcs} task.
  The resultant astrometry of the image improved significantly over the
 original Elixir astrometry solution.

  Ultimately, we used SDSS sources with $16<u<22$ mag and $17<g<22$ mag
 to calculate the astrometric solutions for the $u$- and $g$-band images,
 respectively. The astrometric calibration objects were inspected by
 eye in order to exclude heavily saturated sources.
  The sources used for the astrometric calibration are not
 restricted to stars, since the SDSS filters and CFHT $u^{*}$ and
 $g$ filters are similar enough that the centroiding problem is not
 a great concern. Including both stars and galaxies, we could get
 $\sim$90 objects per chip to derive the astrometric solution. 
  The solutions are calculated interactively for each chip, therefore
 there are 36 solutions for each dither image. 
  Functions used for the fitting are polynomials of order 3$\sim$4,
 and the fitting errors are typically below 0$\farcs$1. 
  How the astrometric solutions fair with other data are described
 in a separate section (\S5.1). 
  Figure {\ref{fig:astrom_imp}} demonstrates the improvement of
 astrometric solutions after executing the above procedure.

  \subsection{Production of Mosaic Images}

       \begin{figure*}
     \begin{center}
     \includegraphics[width=140mm]{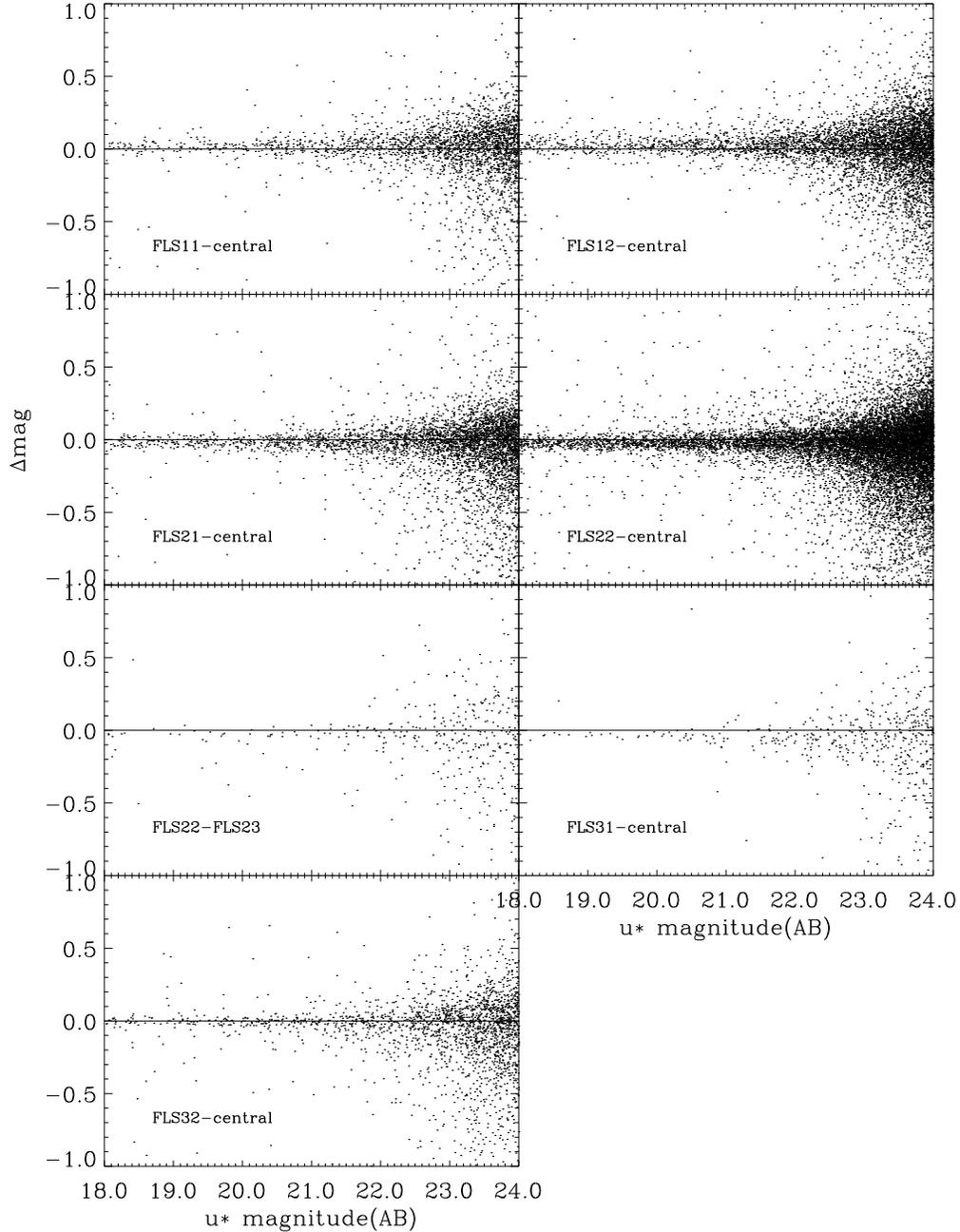}
     \caption{\label{fig:zp_check}  The magnitude difference of
     matched objects from different 8 $u^*$-band fields before
     zeropoint correction is shown. Note that the objects lying
     in FLS21, FLS22, and FLS31 fields are brighter than the
     objects lying in the central field before zeropoint correction.
     The mean value of the magnitude offsets are  
     (FLS11$-$central)$\simeq$0.006, (FLS12$-$central)$\simeq$0.018,
     (FLS21$-$central)$\simeq$$-$0.028, (FLS22$-$central)$\simeq$$-$0.029,
     (FLS22$-$FLS23)$\simeq$$-$0.009, (FLS31$-$central)$\simeq$$-$0.035, and
     (FLS32$-$central)$\simeq$$-$0.001. Since the magnitude offsets of
     FLS21, FLS22, and FLS31 fields are consistent within 
     0.01 magnitude, we used these fields as zeropoint reference fields.}
     \end{center}
     \end{figure*} 
 
     \begin{figure*}
     \begin{center}
     \includegraphics[width=140mm]{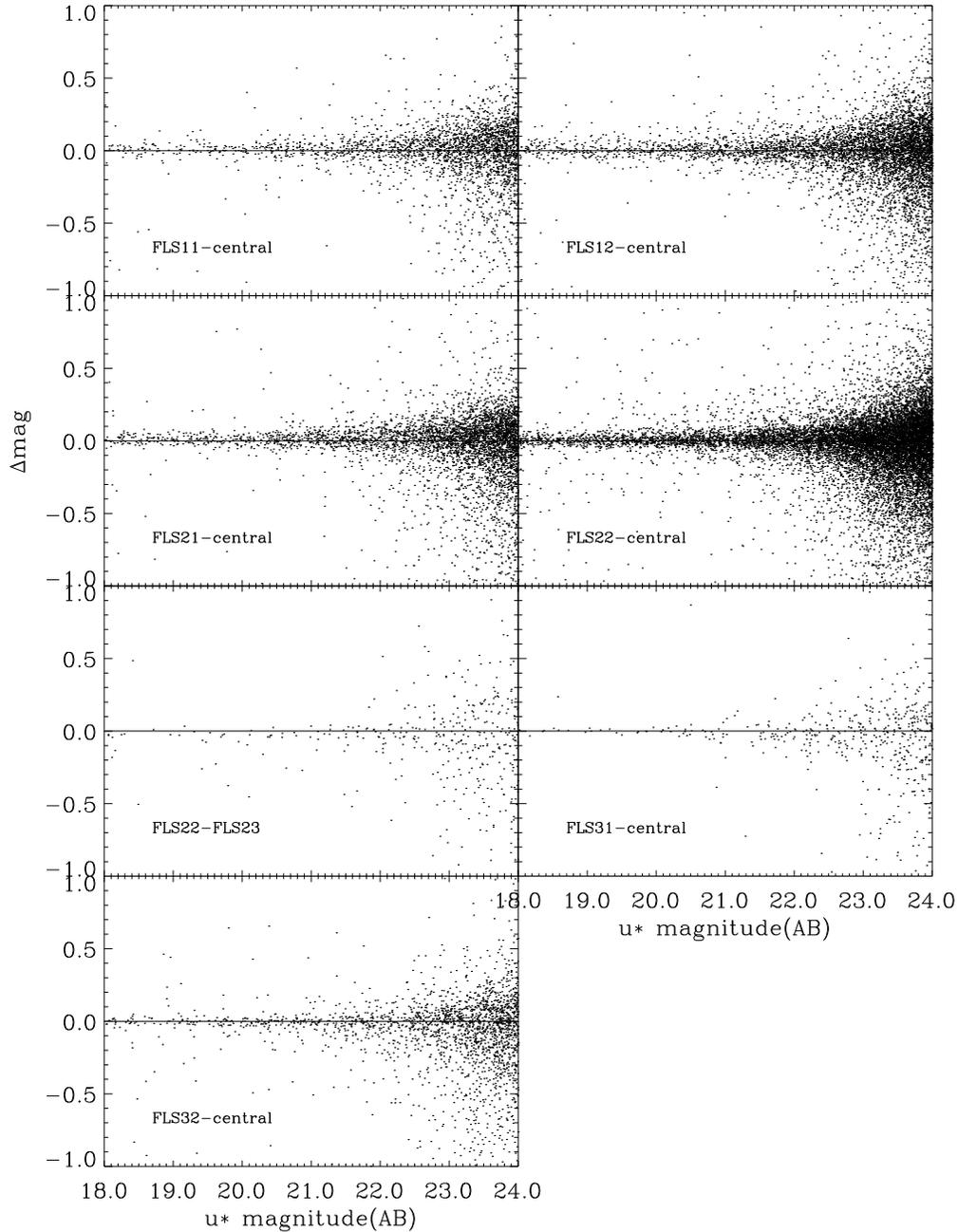}
     \caption{\label{fig:zp_checknew}  After the zeropoint correction,
     the magnitude differences between fields are in the order of
     0.01 magnitude.} 
     \end{center}
     \end{figure*}

  We stacked the post-Elixir processed, astrometrically calibrated
 images using \textit{Swarp} software by E. Bertin at the TERAPIX
 (Traitment Elementaire, Reduction et Analyse des PIXels de
 megacam)\footnote{http://terapix.iap.fr}. 
  By running \textit{Swarp}, we created the mosaic image
 and the corresponding weight image (coverage map) whose pixel
 value is proportional to the exposure time. 
  The reduced images were resampled, background subtracted, flux
 scaled, and finally coadded to one image. The interpolation
 function we used in the resampling is \texttt{LANCZOS3}, which
 uses a moderately large kernel. 
  The pixel scale of the resampled image is kept to the original
 MegaCam pixel size of 0$\farcs$185. The astrometry transformation
 is handled by the \textit{Swarp} program automatically, although
 the \texttt{TNX} WCS as derived in the previous section is not
 a FITS standard. 

  For the background subtraction, we used the background mesh size
 of 128 $\times$ 128 pixels, and applied 3 $\times$ 3 filter box.
  The background mesh size and the bin size are chosen to balance
 out the effect of bright stars (in general smaller than
 128 $\times$ 128 pixels) and the effect of the overall background
 gradient.
  We tried many background combinations to obtain the background
 as flat as possible and settled on the above parameter values.
  The background-subtracted images are combined by taking the
 weighted average of each frame. The weight images are constructed
 by taking the bad pixel masks, and replacing the bad pixel values
 to be 0, and the good pixel values to be 1. The weight images
 constructed this way do not account for small variation in the
 pixel response over the CCD (flat-field).
  Finally, the images are coadded using the following equation:
  \[ F={\frac{\sum w_{i} p_{i} f_{i}}{\sum w_{i}}} \]
  In this equation, $F$ is the value of a pixel in final coadded
 image, $w_i$ is the weight value for the pixel, and $f_i$ is the
 value of the pixel in each individual image. The factor $p_i$
 represents the flux scale for each individual image. 
  The images are calibrated to have the same photometric zeropoints
 after Elixir pipeline, but due to the airmass differences, the
 flux scales of the images are slightly different. 
  In order to take into account the difference in zeropoints, we
 put the flux scale of the image having the least extinction
 to be 1.0.
  The flux scales of other images were calculated according to
 the equation: 
 \[  p_i=10^{\frac{k}{2.5}\times(X_i-X_0)}, \]
 where $k$ is the coefficient for airmass term and $X_0$ and
 $X_{i}$ are airmass values of the reference image and the image
 in consideration, respectively. 
  The coadded gain of the final stacked image varies with 
 position and is proportional to the number of frames used for
 producing each coadded pixel value.

\subsection{Photometric Calibration}

  We performed photometric calibration on the mosaic image of
 each field using the photometric solution provided by the Elixir
 pipeline. 
  During the production of mosaic images (\S3.4), we rescaled
 each dither image so that each coadded image has the photometric
 zeropoint of the image with the least airmass value. 
  Then, these magnitudes were corrected for the galactic extinction
 estimating the amount of galactic extinction from the extinction
 map of Schlegel et al (1998). Since the FLS field lies at
 moderately high galactic latitude, the amount of galactic
 extinction is relatively small. They are 0.15 mag and 0.1 mag on
 average for the $u^*$ and $g$-band respectively.
  In the next step, we derived an additional zeropoint correction
 necessary to account for some of the data that were taken under
 non-photometric condition (e.g., thin cirrus).

  To do so in $u^*$-band, we used fields whose mosaic image is made
 of photometric dither images only. The fields, FLS21, FLS22, and
 FLS31 are such fields, and we used them as reference photometry
 fields to derive photometric zeropoint correction of other
 $u^*$-band fields.
  Figure {\ref{fig:zp_check}} shows the comparison of $u^*$-band
 magnitudes (before correcting for the non-photometric data)
 between different fields that overlap with each other.
  For the comparison of magnitudes, we used the total magnitude
 (\texttt{MAG\_AUTO}) from \textit{SExtractor} (See Section \S4).
  In Figure {\ref{fig:zp_check}}, we can see that $u^*$-band
 magnitudes of objects in FLS21, FLS22, and FLS31 fields are
 slightly brighter than the same objects in the central field. 
  Also, objects in fields other than the reference photometry
 fields have $u^*$-band magnitudes similar or fainter than those
 of the central field. The $u^*$-magnitude offsets between the
 central field and FLS21/22/31 are 0.03 magnitude and the rms in
 the offset values is of order of $\lesssim 0.01$ magnitude.
  This confirms that the u-band magnitudes of FLS21/22/31 fields
 are the most reliable. The final $u^*$-band zeropoint of each 
 field are calculated using this strategy, and we present the
 zeropoints in Table \ref{tab:obslog}. 
  The comparison of $u^*$-band magnitudes of overlapping objects
 after the photometric calibration are presented in Figure
 {\ref{fig:zp_checknew}}.  
  Based on the photometric consistency between FLS21/22/31, we
 estimate the accuracy of the $u^*$-band photometric zeropoint
 to be of order of 0.01 magnitude.

    \begin{figure}
    \begin{center}
    \includegraphics[width=90mm]{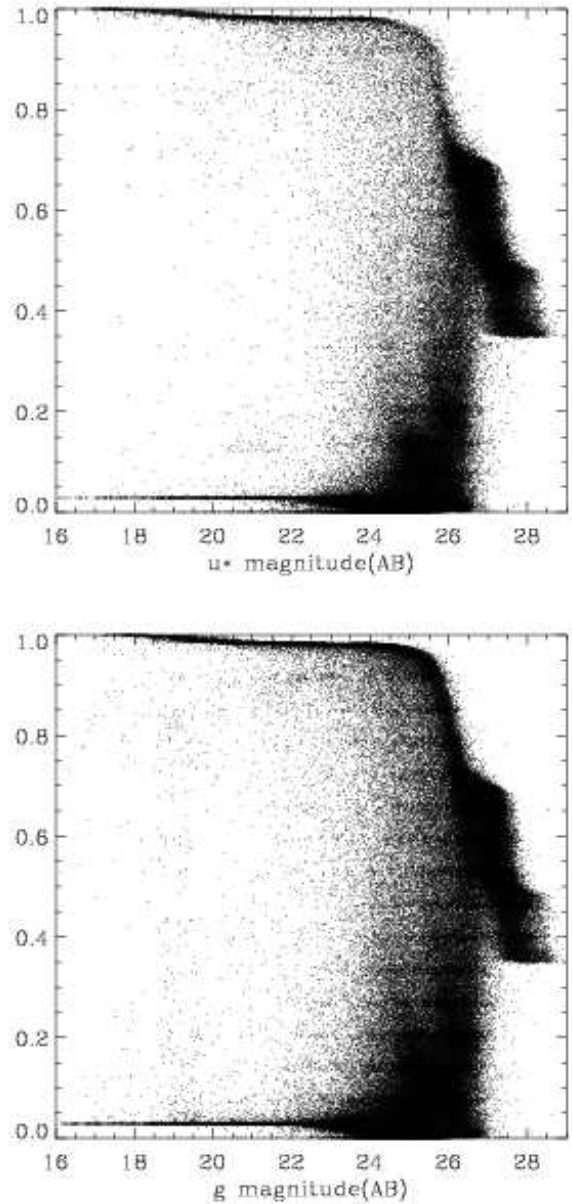}
    \caption{\label{fig:star} The distribution of the values of
    \texttt{CLASS\_STAR} parameter as a function of magnitude
    are presented. At magnitudes below flux limit, the stellarity
    index of the objects are distributed between 0.35 and 0.7.
    The feature of the \texttt{CLASS\_STAR} distribution at faint
    objects ($\le$ 27 mag) is due to the difficulties when
    calculating stellarity index with isophotal areas of the objects.}
    \end{center}
    \end{figure}

  For $g^*$-band mosaic, we preformed photometry on dither images
 that are taken under photometric condition, and derived a necessary
 zeropoint correction by comparing the photometry of the mosaic
 image and the photometric reference image. The $g$-band zeropoint
 of the central field derived this way is also given in
 Table \ref{tab:obslog}.

 \begin{deluxetable*}{ c c c c c c c c c c c c c c }
  \tabletypesize{\scriptsize}
  \tablecaption{Selected 15 entries of $u^{*}$/$g$ band source catalog \label{tab:catalog}}
  \tablewidth{0pt}
  \tablehead{  
  \colhead{obj ID} & \colhead{RA(J2000)} & \colhead{DEC(J2000)} &
  \colhead{u$^*_{AUTO}$\tablenotemark{a}} & \colhead{$\epsilon$} & 
  \colhead{u$^*_{APER}$\tablenotemark{b}} &
  \colhead{$\epsilon$} & \colhead{g$_{AUTO}$} & \colhead{$\epsilon$} &
  \colhead{g$_{APER}$} & \colhead{$\epsilon$} & 
  \colhead{star$_g$\tablenotemark{c}} & \colhead{ext$_{u^*}$\tablenotemark{d}} &
  \colhead{ext$_g$}
  }
\startdata
    1001 & 17:14:13.301 & 59:15:19.199 & 24.992 & 0.051 & 25.016 & 0.042 & 24.726 & 0.066 & 24.699 & 0.052 & 0.010 & 0.120 & 0.088 \\
    1002 & 17:17:11.895 & 59:15:27.342 & 24.882 & 0.043 & 25.200 & 0.052 & 24.834 & 0.063 & 25.112 & 0.073 & 0.006 & 0.169 & 0.125 \\
    1003 & 17:15:52.126 & 59:15:26.820 & 24.872 & 0.040 & 25.184 & 0.048 & 24.796 & 0.053 & 25.135 & 0.064 & 0.086 & 0.151 & 0.111 \\
    1004 & 17:13:19.966 & 59:15:16.411 & 24.955 & 0.050 & 25.408 & 0.061 & 24.582 & 0.069 & 25.144 & 0.092 & 0.024 & 0.114 & 0.084 \\ 
    1005 & 17:19:55.115 & 59:15:21.739 & 25.356 & 0.080 & 25.615 & 0.070 & 25.364 & 0.064 & 25.744 & 0.063 & 0.928 & 0.141 & 0.104 \\
    1006 & 17:14:53.745 & 59:15:25.502 & 25.628 & 0.088 & 26.223 & 0.156 & 25.666 & 0.071 & 26.027 & 0.102 & 0.586 & 0.124 & 0.092 \\
    1007 & 17:13:55.459 & 59:15:21.272 & 25.644 & 0.091 & 26.233 & 0.123 & 25.627 & 0.072 & 26.334 & 0.109 & 0.232 & 0.118 & 0.087 \\
    1008 & 17:18:35.120 & 59:15:27.905 & 26.073 & 0.138 & 27.508 & 0.390 & 26.164 & 0.110 & 28.560 & 0.752 & 0.725 & 0.158 & 0.116 \\
    1009 & 17:15:37.141 & 59:15:28.592 & 26.565 & 0.189 & 26.240 & 0.122 & 26.578 & 0.181 & 26.348 & 0.128 & 0.693 & 0.146 & 0.107 \\
    1010 & 17:15:33.977 & 59:15:19.844 & 23.060 & 0.009 & 24.161 & 0.019 & 22.738 & 0.013 & 23.741 & 0.026 & 0.028 & 0.145 & 0.106 \\
    1011 & 17:14:33.853 & 59:15:19.858 & 23.000 & 0.008 & 23.381 & 0.010 & 22.917 & 0.011 & 23.295 & 0.013 & 0.048 & 0.121 & 0.089 \\
    1012 & 17:17:52.075 & 59:15:25.035 & 23.869 & 0.018 & 24.589 & 0.030 & 23.858 & 0.037 & 24.674 & 0.068 & 0.029 & 0.180 & 0.132 \\
    1013 & 17:18:54.719 & 59:15:23.401 & 23.485 & 0.013 & 23.992 & 0.016 & 23.444 & 0.015 & 23.960 & 0.019 & 0.959 & 0.148 & 0.109 \\
    1014 & 17:16:11.917 & 59:15:26.175 & 23.960 & 0.018 & 24.495 & 0.025 & 23.890 & 0.023 & 24.382 & 0.031 & 0.036 & 0.153 & 0.113 \\
    1015 & 17:17:19.797 & 59:15:29.375 & 26.003 & 0.113 & 26.299 & 0.139 & 26.031 & 0.169 & 26.255 & 0.194 & 0.131 & 0.172 & 0.127 \\
 \enddata
 \tablenotetext{a}{The \texttt{AUTO} magnitudes are calculated by \textit{SExtractor}, using Kron-like elliptical aperture (Kron 1980).}
 \tablenotetext{b}{The \texttt{APER} magnitudes are calculated over a circle with 3$\arcsec$ diameter.}
 \tablenotetext{c}{The stellarity represents \texttt{CLASS\_STAR} calculated with \textit{SExtractor}. The values are distributed between 0 (galaxy) and 1 (star).}
 \tablenotetext{d}{The galactic extinction values are calculated using the extinction map of Schlegel et al.(1998)}
  \end{deluxetable*}

  \section{CATALOGS}

  \subsection{Object Detection and Photometry}

  In the central 1 deg$^2$ region (the 03B run), the stacked
 $u^{*}$- and $g$-band images were registered to the $g$-band
 image. 
  Both bands have the same pixel scale, so they can be easily
 registered with the \texttt{xregister} task in IRAF. 
  After the registration,  source catalogs were created using
 dual mode photometry with \textit{SExtractor}
 (Bertin \& Arnouts 1996).
  Specifically, objects are detected in the $g$-band image, and
 photometry is performed on both the $g$- and $u^*$-band images
 based on these positions.
  The weight image (coverage map) produced by \textit{Swarp}
 was used as the \texttt{WEIGHT\_IMAGE}.  
  The \texttt{LOCAL} background is estimated using a
 128 $\times$ 128 pixel background mesh with 3 $\times$ 3 median
 boxcar filtering, identical to the default \textit{Swarp} settings.

   We filter our images with a gaussian convolution kernel
 (\texttt{gauss\_4.0\_7$\times$7.conv}) matched to our average
 seeing conditions (FWHM$\sim$$0\farcs85$).  Finally, source
 detection was performed on the background-subtracted, filtered
 image by looking for pixel groups above the detection threshold.
  After playing with various parameter combinations and inspecting
 the results by eye, we settled on a minimum of 8 connected pixels
 and a 1.2$\sigma$ minimum detection threshold.
  We also tried various values for the deblending parameter to
 separate objects that are close together, and we adopted a
 deblending threshold of 32 and a deblending minimal contrast
 of 0.005.

  Using dual mode photometry, we detect $\sim$200,000 sources
 in the central 1 deg$^2$ field. We also performed a single mode
 photometry on the $u^*$-band image with the same configuration
 file, to include $u$-band objects that are not detected in $g$-band.
  About $\sim$150 objects are detected in $u^*$-band without
 $g$-band detection.

  We applied a nearly identical method and parameters for
 source detection on the seven 04A FLS $u^*$-band images.
 Since there are several very bright stars in the 04A
 $u^*$-band images, we chose a bigger background mesh size
 -- 256 $\times$ 256 pixels to avoid the over-subtraction
 of the background near bright stars.
  Since the seeing during 04A was slightly worse than that
 of 03B (Table {\ref{tab:obslog}}), we adopted a  slightly
 larger gaussian convolution kernel
 (\texttt{gauss\_5.0\_9$\times$9.conv}).
  As stated earlier, the gain for each stacked image is
 determined to be the original gain times the number of stacked
 images.

  Due to interchip gaps and our 30$\sim$40$\arcsec$ dither
 steps, the effective exposure time varies from pixel-to-pixel.
  We therefore applied the \textit{Swarp} weight image when
 performing photometry.
  Through the visual inspection of the weight image, we
 determined the value of \texttt{WEIGHT\_THRESH} parameter
 for weighting in \textit{SExtractor}. The value of the pixels
 in interchip gaps are measured, and used as threshold parameter
 value.
  This procedure prevents the rise of unusual sequence
 in the magnitude vs. magnitude error plot.

  In the catalogs, we present the apparent magnitude of the
 objects in two different ways (aperture \& total magnitude).
 We measured the aperture magnitude using 3$\arcsec$ diameter
 apertures.
  We adopted the auto magnitude (\texttt{MAG\_AUTO}) from
 \textit{SExtractor} as the total magnitude, which is calculated
 using the Kron elliptical aperture (Kron 1980). The related
 parameters, \texttt{PHOT\_AUTOPARAMS} were determined to be 2.5,
 and 3.5, which are the values in default configuration file.

    \begin{figure}
    \epsscale{1.2}
    \plotone{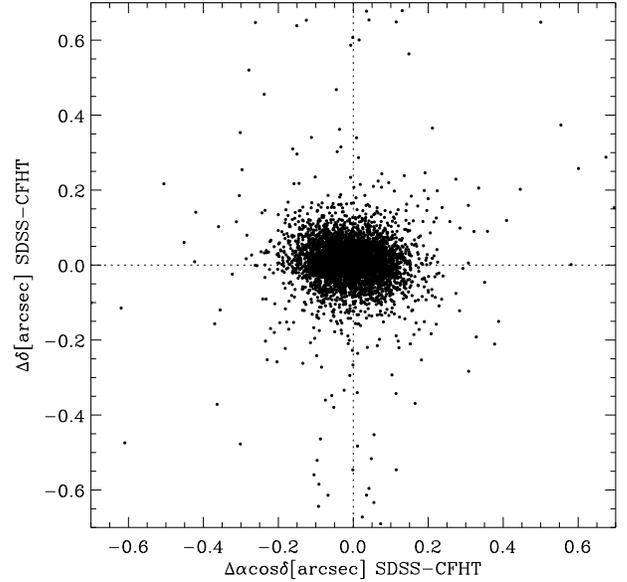}
    \caption{\label{fig:astrom} Comparison between the positions
    of matched objects in our $g$-band catalogs and SDSS $g$-band
    catalogs.
    Offsets are computed as SDSS minus CFHT positions.}
    \end{figure}

    \begin{figure*}
    \plotone{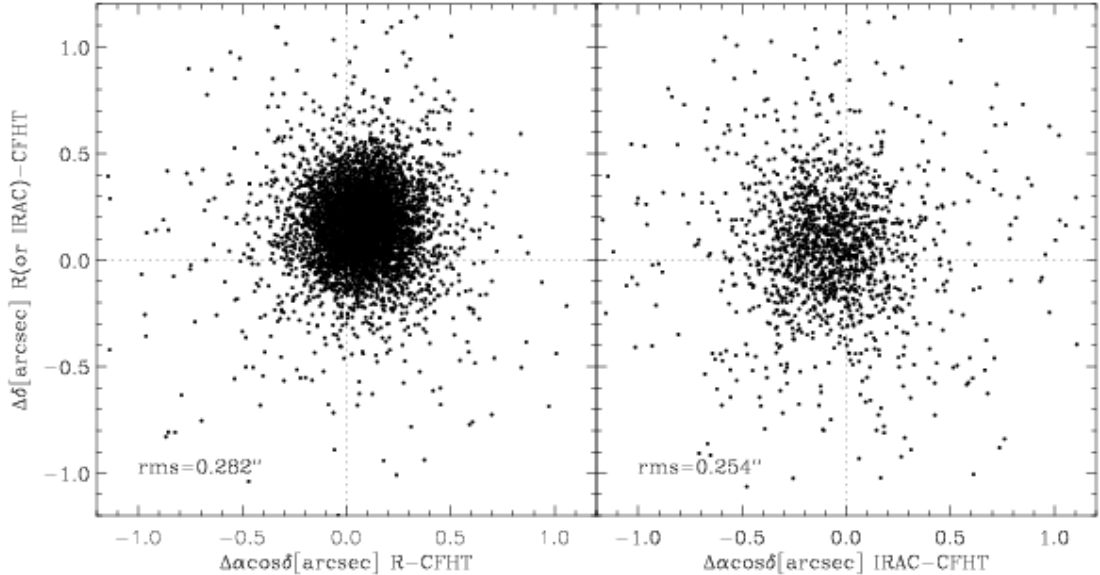}
    \caption{\label{fig:astrom_anc}
     Comparison between the positions of matched objects in our
     $u^*$-band catalogs and $R$-band, IRAC catalogs. The offsets
     are computed as other survey minus CFHT.}
    \end{figure*}

   \subsection{Star--Galaxy Separation}

  We use the \textit{SExtractor} stellarity index to separate
 star-like objects from extended sources.
  \textit{SExtractor} uses 8 measured isophotal areas, peak
 intensity and seeing information to calculate the stellarity
 index.
  With the \texttt{SEEING\_FWHM} parameter value and the neural
 network file as the inputs, \textit{SExtractor} gives the
 stellarity index (\texttt{CLASS\_STAR}), which has values
 between 1 (starlike object) and 0 as the result of the classification.
  Figure {\ref{fig:star}} shows the distribution of stellarity
 index as a function of magnitude in the central 1 deg$^2$ field.
  We randomly picked objects with different apparent magnitudes
 and stellarity index values, and visually inspected them to see
 how useful the stellartiy index is for the point source vs.
 extended source classification.
  Our visual inspection reveals that bright objects ($g<21$ mag,
 and $u^{*}<20$ mag) with the stellarity index $\gtrsim 0.8$ are
 found to be stars.
  Through the visual inspection, we adopted the following criteria
 for separating point sources from extended objects:
\begin{center} 
\texttt{CLASS\_STAR}$>$0.8 for 20$<u<$23 \\  
\texttt{CLASS\_STAR}$>$0.8 for 21$<g<$23 \\
  \end{center}

   In addition to these criteria, we weed out bright saturated
 stars ($g<21$ mag, $u^{*}<20$ mag), based on SDSS point source
 catalogs. The subtraction of objects matched with SDSS stars is
 a good method for removing stars from our catalogs.

  On the other hand, it is difficult to address the nature of
 faint ($g>23$ mag, $u^*>23$ mag) objects with the stellarity
 value alone, because faint, distant small galaxies are hardly
 resolved under 1$\arcsec$ seeing conditions.
  Fortunately, number counts tend to be dominated by galaxies
 at $u^{*}, g > 23$ mag. Therefore we do not attempt to separate
 stars from galaxies at these flux levels.
  Also, the stellarity index distribution shows a converging
 feature between 0.35 and 0.7 at magnitudes below flux limit
 ($g>26.5$ mag, $u^*>26.2$ mag). This is thought to be the
 effect of difficulties in determining isophotal areas of
 faint objects.

\subsection{Catalog Formats}

  As an example, 15 entries of the central 1 deg$^2$ $u^{*}$- and
 $g$- band merged catalog is presented in Table {\ref{tab:catalog}}.
  All magnitudes are given in the AB magnitude system.
  We include the total magnitude (\texttt{MAG\_AUTO}) and the
 aperture magnitude (3$\arcsec$ diameter) in the catalog.
  The magnitudes are not corrected for the galactic extinction,
 but the extinction values are presented as a separate column.
  Magnitude errors are the outputs from \textit{SExtractor}.
  Considering the error in zeropoint and the extinction correction,
 we believe that the minimal error in the magnitude will be about
 0.02 mag.
  Stellarity indices are calculated in the $g$-band image for
 the central field.

\section{PROPERTIES OF THE DATA}

  \subsection{Astrometry}

  We derived the final astrometric solution using SDSS sources
 (both point and extended sources).
  The rms error between SDSS point sources and our sources are
 calculated to be 0$\farcs$1$<dr<$0$\farcs$15. Considering the
 absolute astrometric error of 0$\farcs$1 in SDSS (Pier et al.
 2003), our catalogs are thought to have the positional error of
 roughly 0$\farcs$15 -- 0$\farcs$2.
  Figure {\ref{fig:astrom}} shows the positional differences
 between SDSS sources and our sources over the entire mosaic of
 the central 1 deg$^2$ field.

  To address the accuracy of astrometric calibration in other ways,
 we compared our source positions with other available catalogs
 for the \textit{Spitzer} FLS.
  The catalogs compared are KPNO $R$-band data (Fadda et al. 2004)
 and the \textit{Spitzer} IRAC data (Lacy et al. 2005).
  The results are given in  Figure {\ref{fig:astrom_anc}}.
  The average rms error,
 $dr=\sqrt{(\Delta \alpha cos \delta ^{2} + \Delta \delta ^{2})}$,
 is about $\sim$0$\farcs$28 with respect to $R$-band data. 
  $R$-band positions have a systematic offset of
 $\Delta\alpha=0\farcs09\pm0\farcs27$ and
 $\Delta\delta=0\farcs19\pm0\farcs31$ with respect to CFHT positions.
  The systematic offset is originated from the fact that the
 reference catalog for $R$-band data is GSC II stars, which is
 known to have a systematic offset with respect to SDSS reference
 positions (Fadda et al. 2004). Therefore, we consider the 
 astrometry difference between $R$-band and CFHT data is well
 understood.
  For IRAC sources, the average rms error is $dr\simeq$ 0$\farcs$25.
  The offset of IRAC positions according to CFHT positions are
 calculated to be
 $\Delta\alpha=-0\farcs08\pm0\farcs40$ and
 $\Delta\delta=0\farcs04\pm0\farcs38$.
  The FLS IRAC data have positional accuracy of 0$\farcs$25 rms
 with respect to 2MASS (Lacy et al. 2005). The difference between
 2MASS positions and SDSS positions caused the offset between 
 IRAC positions and CFHT positions. The results add support to
 the astrometric accuracy of our dataset.

  \subsection{Photometry Transformation}

\begin{figure}
     \epsscale{1.2}
     \plotone{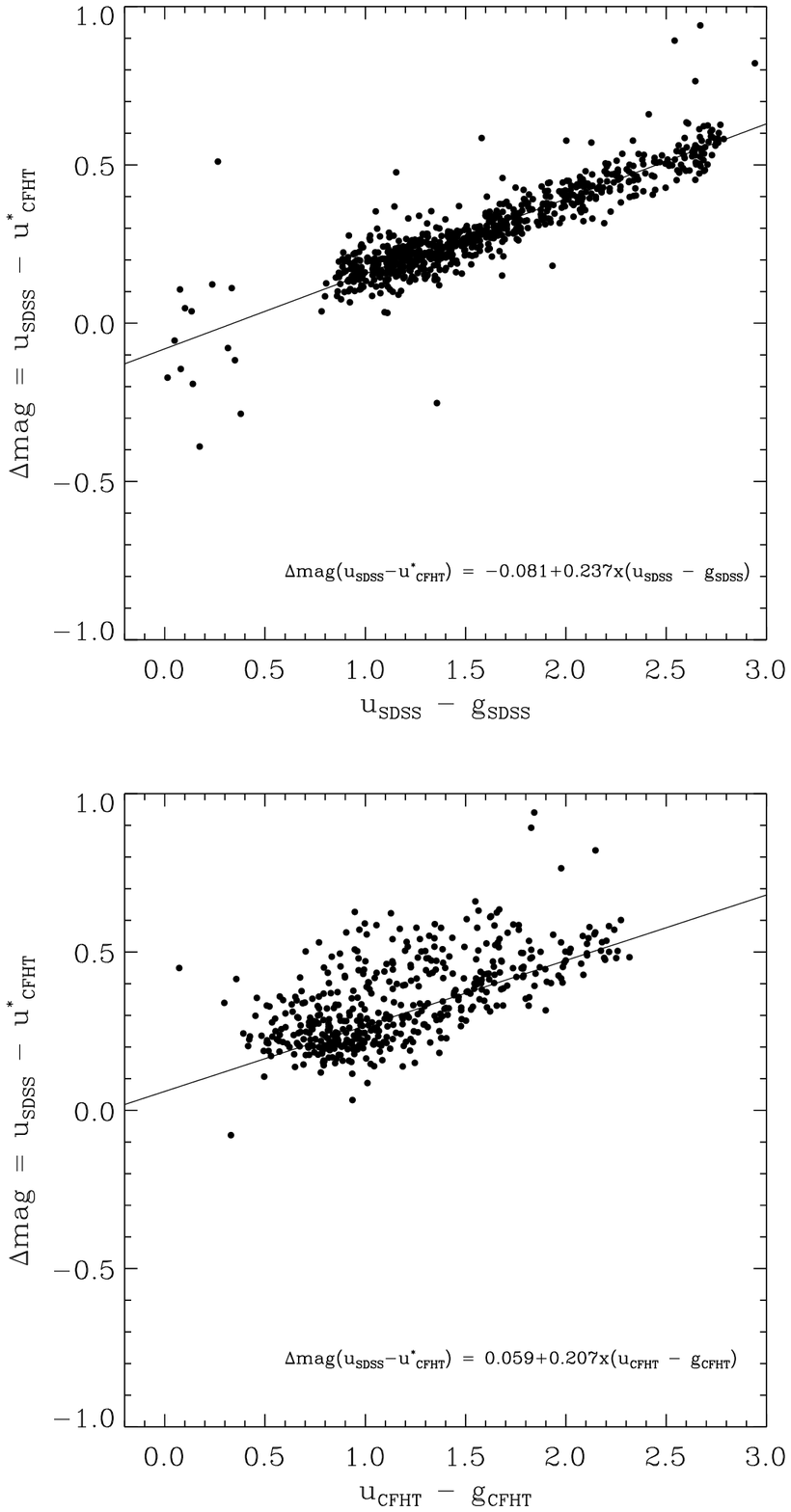}
     \caption{\label{fig:filconv} 
     Empirical difference between CFHT $u^*$-band and
     SDSS $u$-band magnitudes as a function of ($u-g$) color. 
     We used least--square fit to find the conversion equation
     from one $u$-band to another.
     }
     \end{figure}

  As we have shown in Figure {\ref{fig:filter}}, MegaCam filters
 are slightly redder than SDSS counterparts.
  Therefore, we examined the correlations between $u^*$ magnitudes
 and SDSS $u$ magnitudes.
  To do so, we matched non-saturated point sources with 18.5$<u<$20
 from our catalog with the SDSS point source catalog, and compared
 the differences in their magnitudes. 
  Figure {\ref{fig:filconv}} shows the comparison, and it
 demonstrates that the difference between the $u^{*}$ and $u$
 varies from 0.1 to 0.6 magnitude as a function of the color of
 the object. 
  We fitted the relation with a first order polynomial to construct
 the conversion formula between CFHT $u^*$ and SDSS $u$-band
 magnitudes, and obtained the following relation from the linear
 least--square fitting.
  Although there exist some scatter, the conversion formula could
 be a good reference in use of CFHT $u^*$ magnitude.

 \begin{center}
$\Delta mag(u_{SDSS}-u_{CFHT})=(-0.081\pm0.006)+(0.237\pm0.004)\times(u-g)_{SDSS}$
 \end{center}

     \begin{figure}
     \epsscale{1.2}
     \plotone{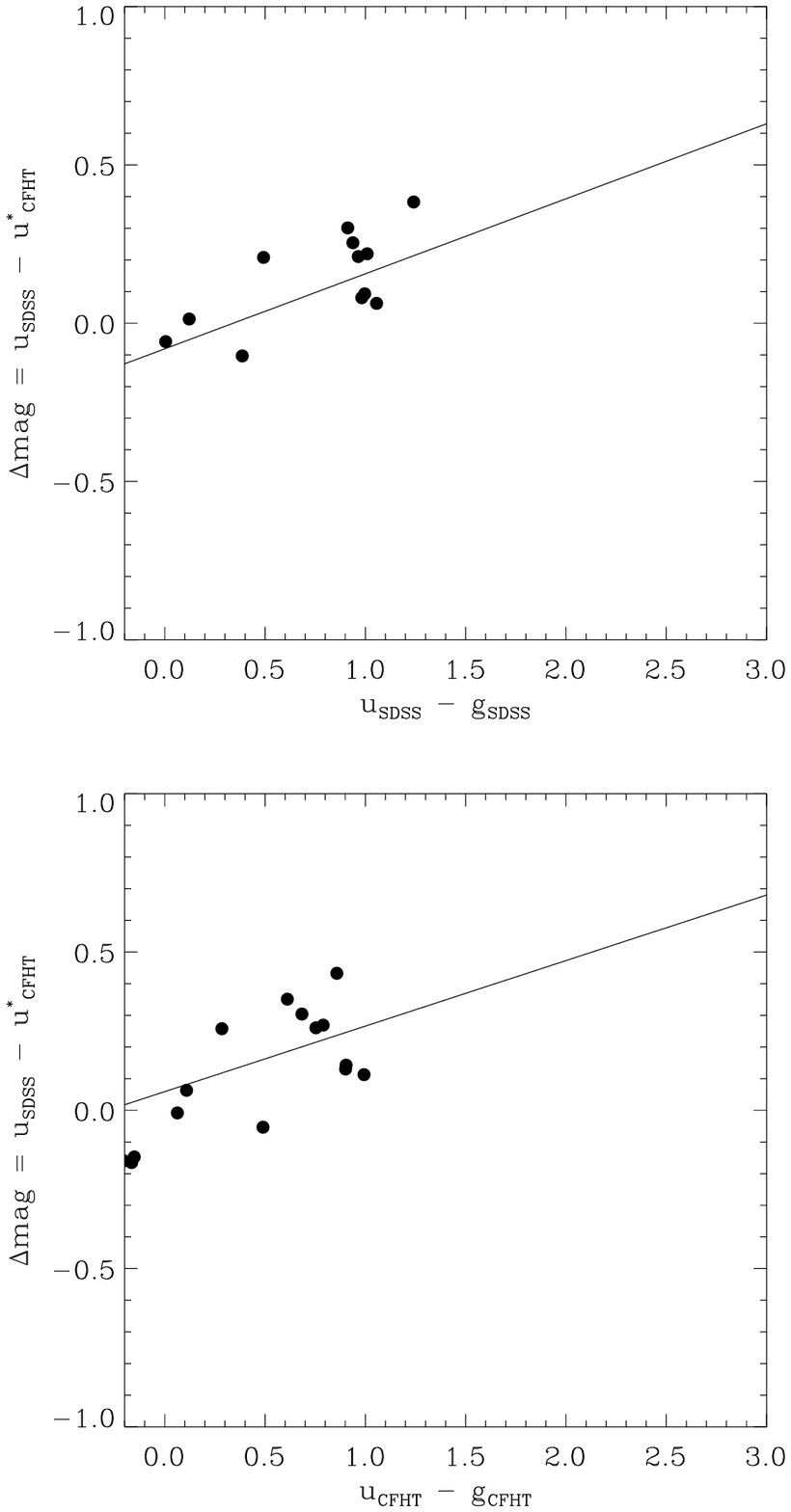}
     \caption{\label{fig:magdf_star} 
     Calculated $u^{*}-u$ for various spectral types of stars. 
     The stellar templates from O5V to F9V stars were used
     in calculation, since in case of red stars, $u$-band is not
     good filter to sample their red colors.
     }
     \end{figure}

     \begin{figure}
     \epsscale{1.2}
     \plotone{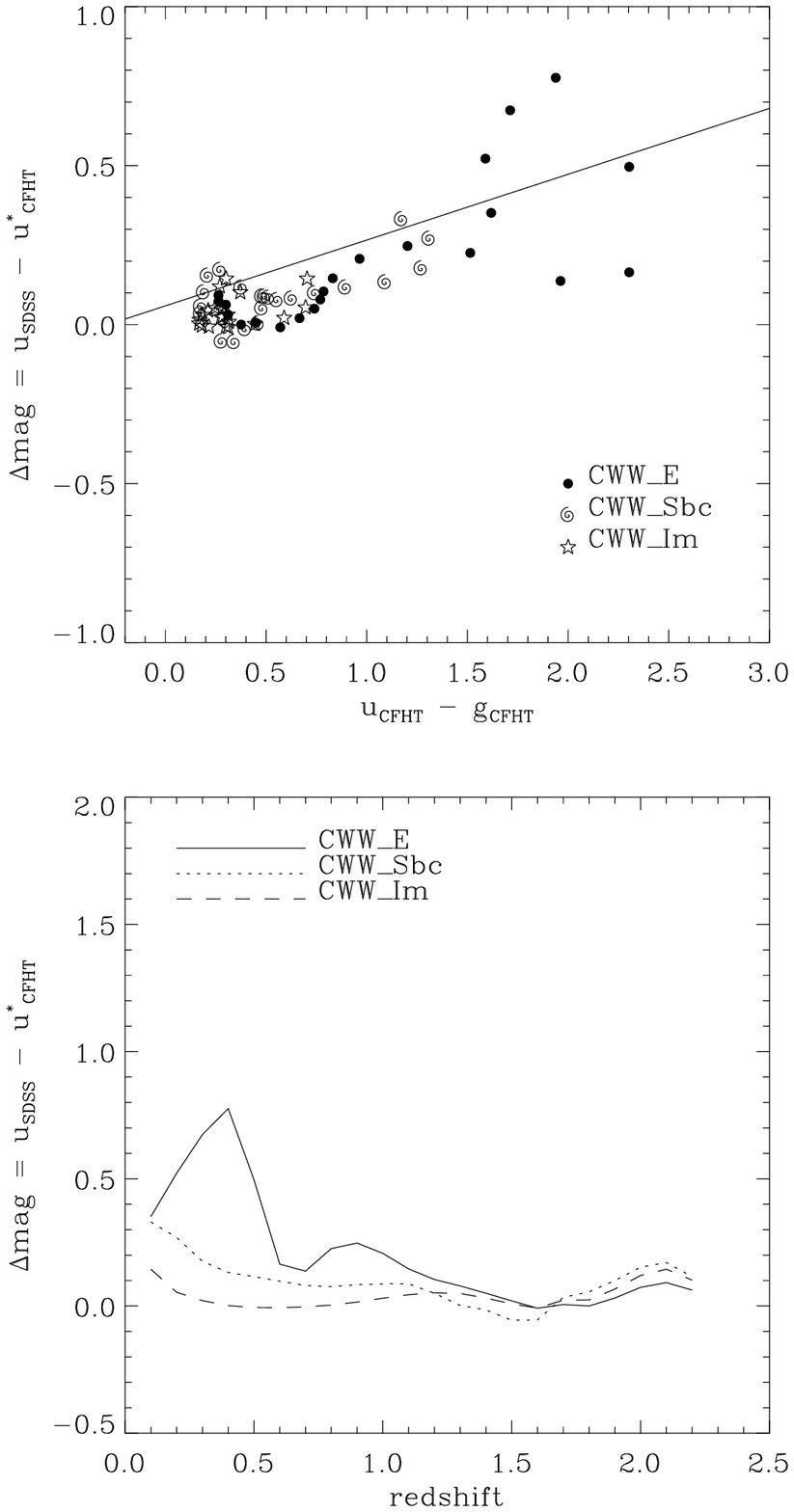}
     \caption{\label{fig:magdf_gal}
     Calculated difference 
     between the CFHT $u^*$-band 
     and SDSS $u$-band magnitudes 
     as a function of ($u - g$) color. The galaxy templates used
     are elliptical, spiral, \& irregulars from Coleman, Wu,
     \& Weedman 1980. We redshifted the galaxies to various
     redshifts, and calculated their colors using filter curves.
     }
     \end{figure}

     \begin{figure}
     \epsscale{1.2}
     \plotone{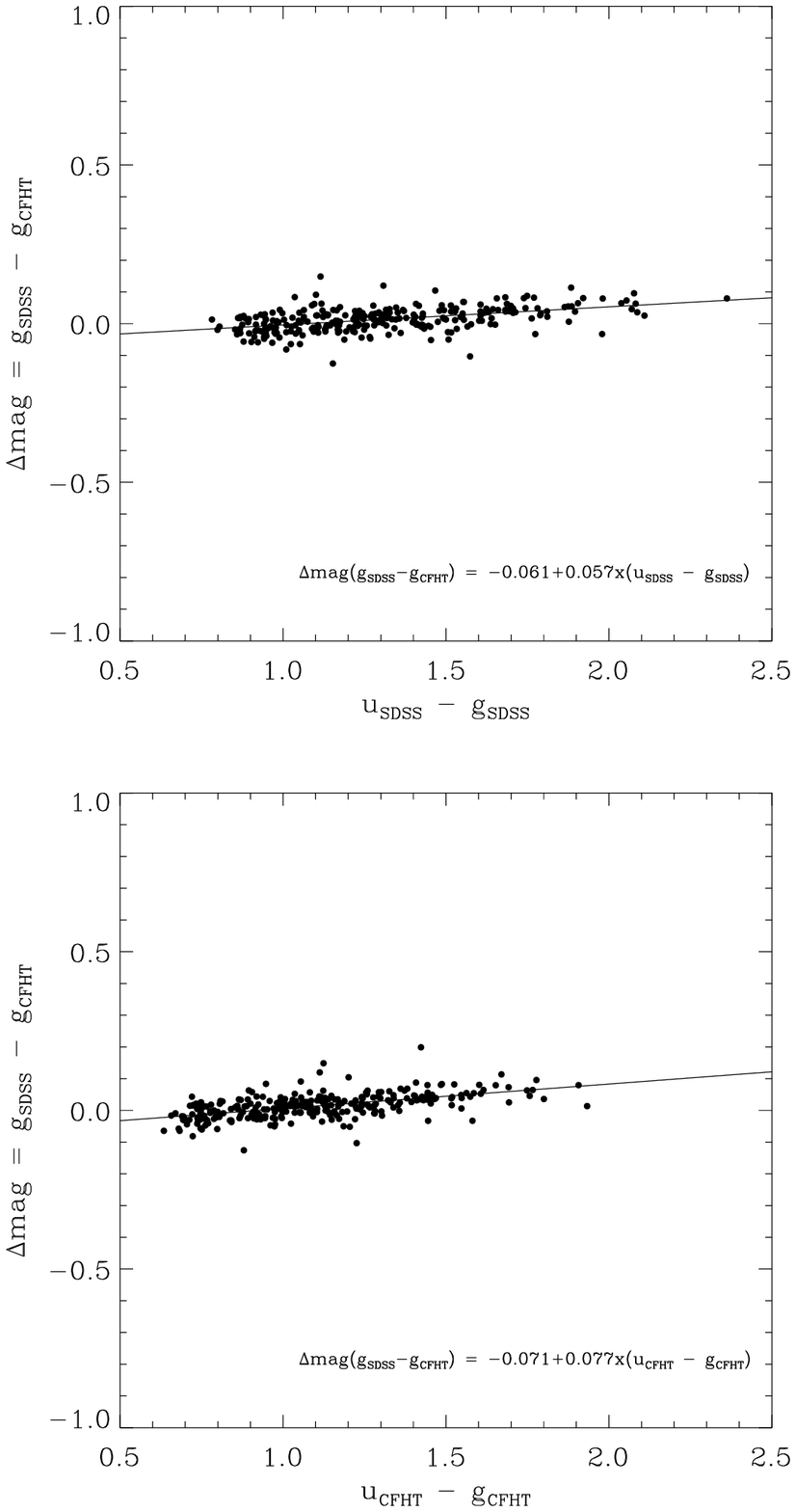}
     \caption{\label{fig:filconv_g}
     Empirical difference of CFHT $g$-band magnitude and SDSS
     $g$-band magnitude according to ($u-g$) color.
     Because CFHT filter is redder than SDSS filter, there is
     also a small tendency of magnitude difference along the color.
     But it is not difficult to say that the difference is negligible.}
     \end{figure}

     \begin{figure}
     \epsscale{1.2}
     \plotone{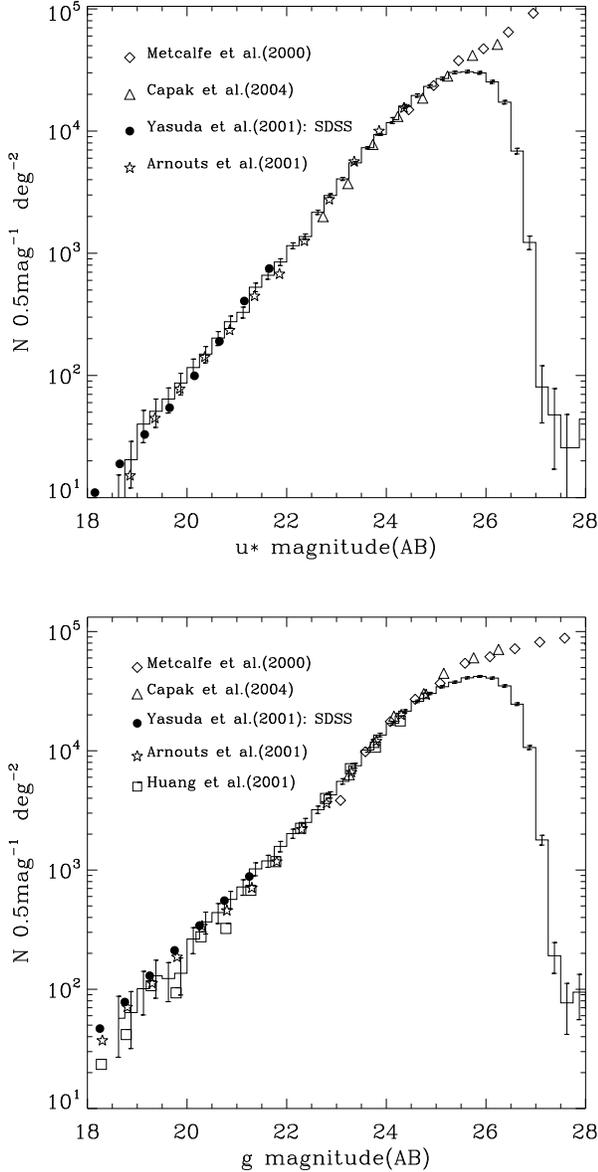}
     \caption{\label{fig:nc}
     Galaxy number counts in FLS region from our $u^*$,
     $g$-band images. Results from other studies are overplotted.
     Magnitudes are in AB system.}
     \end{figure}

     \begin{figure}
     \epsscale{1.2}
     \plotone{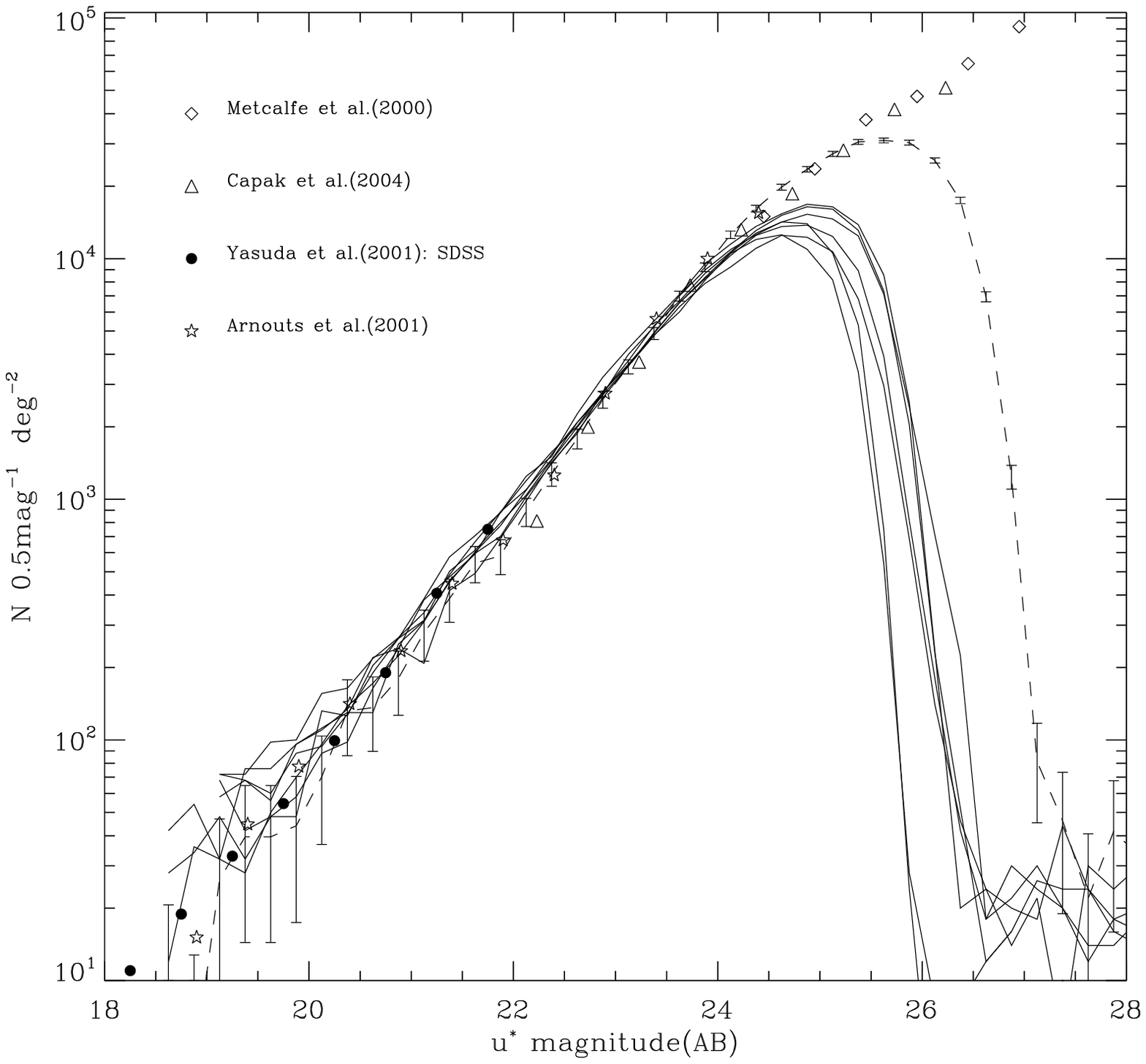}
     \caption{\label{fig:nc_u04}
     Galaxy number counts in FLS region from the shallower
     $u^*$-band images. Different solid lines indicate the
     different fields. The shallowest field, [FLS23] shows
     100\% completeness to about 23.6 magnitude.
     The deepest field, [FLS11] is 100\% complete down to 24.2 mag.
     The dashed line is the result from the central 1 deg$^2$ region.
     }
     \end{figure}

  We checked the above empirical relation between MegaCam
 $u^*$ and SDSS $u$ by calculating ($u^{*}-u$) vs. ($u-g$)
 using the filter curves in Figure {\ref{fig:filter}} and
 the stellar templates from O5V to F9V stars (Silva \& Cornell 1992),
 or the empirical galaxy templates at different redshift
 (Coleman et al. 1980).
  The theoretically calculated relations of ($u^{*}-u$) vs.
 ($u-g$) agree well with the empirical relations (Figure
 {\ref{fig:magdf_star}}, {\ref{fig:magdf_gal}}).

  On the other hand, the difference between SDSS $g$-band
 magnitude and CFHT $g$-band magnitude is considerably small. 
  Figure {\ref{fig:filconv_g}} shows their difference along
 ($u-g$) color.
  Compared to Figure {\ref{fig:filconv}}, the magnitude
 difference is small (they scatters within 0.15 mag), but the  
 differences still have a tendency of increasing toward red color.
  We applied least--square fit to $g$-band magnitude difference
 to find the following conversion relation. 

 \begin{center}
$\Delta mag(g_{SDSS}-g_{CFHT})=(-0.061\pm0.007)+(0.057\pm0.004)\times(u-g)_{SDSS}$
 \end{center}

  \subsection{Galaxy Number Counts}

  To show the depth and homogeneity of the survey, we present
 the galaxy number counts in $u^*$ and $g$-band.
  The number counts are constructed for both the deep central
 1 deg$^2$ field as well as the shallower $u^*$-band coverage
 of the entire FLS fields.
  Then, the results are compared with galaxy counts from other
 studies.
  The direct comparison is possible for the counts constructed
 from SDSS filters (Yasuda et al. 2001), while, for other studies
 using Johnson $U$, we converted the Johnson $U$-band magnitude
 to our $u^*$ magnitude using the $u-U\simeq0.8$ conversion
 relation of galaxy colors from Fukugita et al. (1995), and
 then applying the mean of our empirical conversion relation
 between $u^*$ and $u$ ($u^{*} - u = 0.4$ mag).
  For $g$ magnitude, we used $B$-band number counts from the
 literatures (Metcalfe et al. 2000; Capak et al. 2004;
 Arnouts et al. 2001; Huang et al. 2001), converting $B$ mag to
 $g$ mag according to the relation of $g-B=0.3\sim0.6$
 (Fukugita et al. 1995).
  The conversion relation from Johnson $U$ to $u$, and from
 $B$ to $g$ filter varies according to the shape of the galaxy
 spectral energy distribution, so we adopt the mean difference
 to translate one filter to another.

  The number counts are shown in Figure {\ref{fig:nc}}.
  When constructing the number counts, we excluded stellar
 objects using the stellarity cut introduced in section \S4.2.
 For objects brighter than $g<21$ mag and $u^{*}<20$ mag, we
 matched the SDSS point source catalogs with our catalogs, and
 subtracted matched objects which are considered to be stars.
  Error bars on the figure represent the poisson errors only.
  Also note that our counts are not corrected for completeness.
   
  With the exception of SDSS studies (Yasuda et al. 2001),
 galaxy counts are typically derived from deep, but small area
 surveys ($\sim$0.2--0.3 deg$^2$).
  Our image covers from 0.94 deg$^2$ (in case of $g$-band) to 
 5.13 deg$^2$ (in case of shallower $u^*$-band). 
  For $u^{*}<24$ mag, we include the shallower $u^{*}$-band
 data to maximize the area  coverage. We can say that our number
 counts are better than other number counts in terms of the area
 coverage. 
  For future use, we present a table that summarizes the result
 of galaxy number counts (Table {\ref{tab:numbers}}). To construct
 the number counts, we use the auto magnitude in \textit{SExtractor}.
  Our number counts are in good agreement with the counts from
 other studies to $u^*\simeq24.8$ mag, $g\simeq25.2$ mag.
  Beyond  $u^*\simeq24.8$ mag and $g\simeq25.2$ mag, our data
 starts to tail off, and this shows that our $u^*$ and $g$ catalogs
 are nearly 100\% complete down to the above magnitude limits.
  These limiting magnitudes have the uncertainty of 0.3 magnitudes
 due to the fluctuation in the galaxy number counts at the faint end.
  We also obtained the number counts for the shallower $u^*$-band
 data, the results are illustrated in Figure {\ref{fig:nc_u04}}.
  In case of the deepest field among the shallower $u^*$-band,
 the 100\% limit is 24.2 magnitude. 

  \subsubsection{Completeness} 
  
\begin{figure}
     \epsscale{1.2}
     \plotone{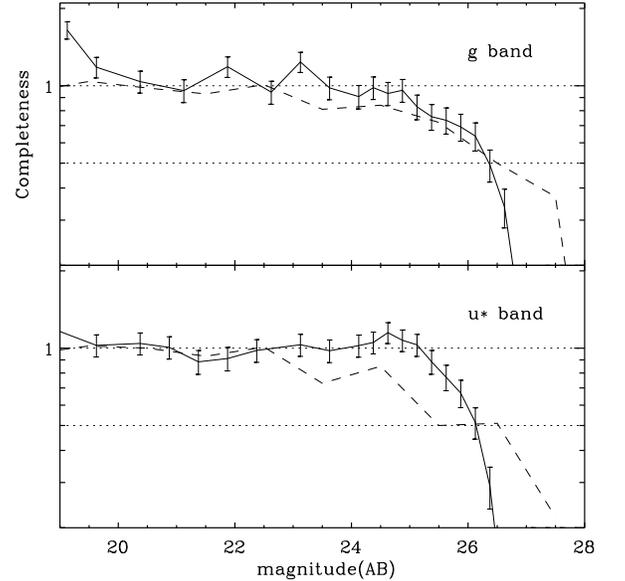}
     \caption{\label{fig:comp} The completeness test. Completeness
     was inspected by dividing galaxy number counts with median
     number counts from other surveys (\textit{solid} line). The
     \textit{dashed} line shows the estimation of completeness
     from artificial image made.}
     \end{figure}

  To inspect the completeness of our survey in more detail,
 we used two independent methods. 
  First, we compared our galaxy counts with the deeper galaxy
 counts from other studies.
  Second, we made artificial images having the same properties
 with our observation (seeing, crowding, gain \& background)
 using \texttt{artdata} package in IRAF. The artificial image
 contained both point sources and extended sources. 
  The extended sources are defined to have 40 \% of whole galaxies
 as elliptical galaxies and the remainders for disk galaxies,
 with minimum redshift of $z\simeq0.01$. Poisson noise is added
 to the background. 
  Then, we performed photometry with the same configuration file
 we used to make catalogs of observed sources. 
  The results are shown in Figure {\ref{fig:comp}}, where the
 solid line and points indicate the completeness from the first
 method, and the dashed line represents the completeness
 estimated from the simulation.
  Figure {\ref{fig:comp}} shows that the data is 100\% complete
 down to $g \simeq25.2$ mag and $u \simeq24.8$ mag, agreeing the
 value found in the previous section, and is $\sim$50\% complete
 to 26.5 magnitude for $g$-band, 26.2 magnitude for $u^*$-band. 

\begin{figure}
     \epsscale{1.2}
     \plotone{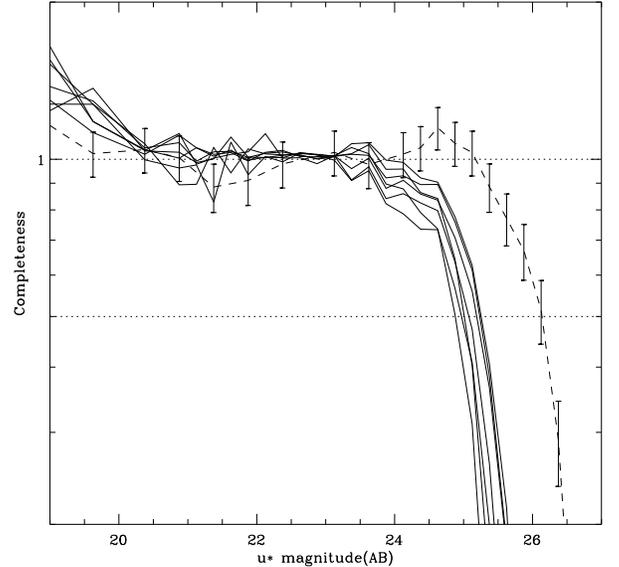}
     \caption{\label{fig:com_u04}
     Completeness of the survey as a function of magnitude for
     the shallower $u^*$-band.
     The dashed line is the completeness of the central 1 deg$^2$
     region $u^*$-band.}
     \end{figure}

  For the shallower $u^*$-band data, we examined the completeness
 using the first method (See Figure {\ref{fig:com_u04}}).
 The 50\% complete limiting magnitudes are 24.8--25.4 
 mag, which are consistent with the estimated value in
 Table {\ref{tab:obslog}}.

  \begin{deluxetable*} {c|ccc|ccc} 
  \tabletypesize{\scriptsize}
  \tablecaption{ The result of galaxy number counts \label{tab:numbers} }
  \tablewidth{0pc}
  \tablehead{
  \colhead{mag (AB)} & \colhead{log $N_{u^*}$ (0.5 mag$^{-1}$ deg$^{-2}$)} &
  \colhead{Number of Galaxies} & \colhead{Area Coverage} & 
  \colhead{log $N_g$} &
  \colhead{Number of Galaxies} &
  \colhead{Area Coverage} \\
  \colhead{ }& \colhead{(0.5 mag$^{-1}$ deg$^{-2}$)} &
  \colhead{Number of Galaxies} & \colhead{(deg$^2$)} &
  \colhead{(0.5 mag$^{-1}$ deg$^{-2}$)} & 
  \colhead{Number of Galaxies} &
  \colhead{(deg$^2$)} 
  }
\startdata 
  18.625 & 0.933 $^{+0.061}_{-0.071}$ & {44} & 5.13 & 1.743 $_{-0.069}^{+0.060}$ & 52 & 0.94 \\ 
  18.875 & 1.272 $_{-0.047}^{+0.042}$ & {96} & 5.13 & 1.790 $_{-0.065}^{+0.057}$ & 58 & 0.94 \\
  19.125 & 1.555 $_{-0.033}^{+0.031}$ & {184} & 5.13 & 1.991 $_{-0.051}^{+0.046}$ & 92 & 0.94 \\
  19.375 & 1.659 $_{-0.029}^{+0.028}$ &  234 & 5.13 & 2.099 $_{-0.045}^{+0.041}$ & 118 & 0.94 \\
  19.625 & 1.760 $_{-0.026}^{+0.025}$ &  295 & 5.13 & 2.076 $_{-0.046}^{+0.042}$ & 112 & 0.94 \\
  19.875 & 1.891 $_{-0.022}^{+0.021}$ &  399 & 5.13 & 2.120 $_{-0.044}^{+0.040}$ & 124 & 0.94 \\
  20.125 & 2.017 $_{-0.019}^{+0.018}$ &  534 & 5.13 & 2.407 $_{-0.031}^{+0.029}$ & 240 & 0.94 \\  
  20.375 & 2.127 $_{-0.017}^{+0.016}$ &  688 & 5.13 & 2.551 $_{-0.026}^{+0.025}$ & 334 & 0.94 \\
  20.625 & 2.258 $_{-0.014}^{+0.014}$ &  930 & 5.13 & 2.629 $_{-0.024}^{+0.023}$ & 400 & 0.94 \\
  20.875 & 2.392 $_{-0.012}^{+0.012}$ & 1266 & 5.13 & 2.740 $_{-0.021}^{+0.020}$ & 516 & 0.94 \\
  21.125 & 2.469 $_{-0.011}^{+0.011}$ & 1512 & 5.13 & 2.844 $_{-0.018}^{+0.018}$ & 656 & 0.94 \\
  21.375 & 2.675 $_{-0.009}^{+0.009}$ & 2426 & 5.13 & 2.995 $_{-0.015}^{+0.015}$ & 930 & 0.94 \\
  21.625 & 2.772 $_{-0.008}^{+0.008}$ & 3038 & 5.13 & 3.062 $_{-0.014}^{+0.014}$ & 1084 & 0.94 \\
  21.875 & 2.881 $_{-0.007}^{+0.007}$ & 3900 & 5.13 & 3.184 $_{-0.012}^{+0.012}$ & 1436 & 0.94 \\
  22.125 & 3.015 $_{-0.006}^{+0.006}$ & 5306 & 5.13 & 3.292 $_{-0.011}^{+0.011}$ & 1840 & 0.94 \\
  22.375 & 3.089 $_{-0.006}^{+0.005}$ & 6302 & 5.13 & 3.385 $_{-0.010}^{+0.010}$ & 2282 & 0.94 \\
  22.625 & 3.288 $_{-0.004}^{+0.004}$ & 9954 & 5.13 & 3.492 $_{-0.009}^{+0.008}$ & 2920 & 0.94 \\
  22.875 & 3.425 $_{-0.004}^{+0.004}$ & 13664 & 5.13 & 3.616 $_{-0.007}^{+0.007}$ & 3884 & 0.94 \\
  23.125 & 3.561 $_{-0.003}^{+0.003}$ & 18673 & 5.13 & 3.729 $_{-0.007}^{+0.006}$ & 5042 & 0.94 \\
  23.375 & 3.693 $_{-0.003}^{+0.003}$ & 25280 & 5.13 & 3.860 $_{-0.006}^{+0.006}$ & 6816 & 0.94 \\
  23.625 & 3.817 $_{-0.002}^{+0.002}$ & 33628 & 5.13 & 3.992 $_{-0.005}^{+0.005}$ & 9224 & 0.94 \\
  23.875 & 3.925 $_{-0.002}^{+0.002}$ & 43196 & 5.13 & 4.118 $_{-0.004}^{+0.004}$ & 12334 & 0.94 \\
  24.125 & 4.021 $_{-0.002}^{+0.002}$ & 53897 & 5.13 & 4.216 $_{-0.004}^{+0.004}$ & 15464 & 0.94 \\
  24.375 & 4.143 $_{-0.004}^{+0.004}$ & 13067 & 0.99 & 4.320 $_{-0.003}^{+0.003}$ & 19652 & 0.94 \\
  24.625 & 4.232 $_{-0.003}^{+0.003}$ & 16055 & 0.99 & 4.411 $_{-0.003}^{+0.003}$ & 23648 & 0.94 \\
  24.875 & 4.307 $_{-0.003}^{+0.003}$ & 19042 & 0.99 & 4.506 $_{-0.003}^{+0.003}$ & 27510 & 0.94 
\enddata
  \end{deluxetable*}
 
     \subsubsection{Reliability}

  We also examined the reliability of our detections.
  We created a negative image by multiplying $-1$ to the mosaic
 image and performed photometry on the negative image. 
  The number of objects detected in the negative image gives us 
 an estimate of how many false detections exist in our catalogs.
  The false detection rate is defined as
  \[ \frac{N_{false}(m, m+\Delta m)}{N(m, m+\Delta m)} \]
 where $N_{false}(m, m+\Delta m)$ is the number of objects
 detected in the negative image over the magnitude bin of
 ($m, m+\Delta m$), and $N(m, m+\Delta m)$ is the number of
 objects detected in the original mosaic over the same
 magnitude bin.
  The false detection rate is 0.5, when there is only noise
 in the image.

  The number of spurious detections are relatively small
 (less than 0.7\%) until $g\simeq26.5$ mag and $u^{*}\simeq26.0$ mag,
 increases steeply toward fainter magnitudes. The presence of
 false detections at brighter magnitudes (which is above the
 limiting magnitude) is due to the effects of streaks around
 bright stars. On the other hand, the majority of false detections
 at fainter magnitudes is generated from the noise. The similar
 tendency is emerging on the shallower $u^*$-band images. The
 false detection rate remains under 1.8\% until the limiting
 magnitudes for the shallower $u^*$-band.

\section{SUMMARY}

  The entire $\sim$5 deg$^2$ region of the \textit{Spitzer}
 First Look Survey area have been observed with the MegaCam on the
 CFHT 3.6 m telescope.
  From the data, we created nine final mosaic images,
 photometrically and astrometrically calibrated. Two of them 
 are the deeper $u^*$, $g$-band images on the central 1 deg$^2$,
 and the remaining seven are the shallower $u^*$-band images
 over the whole 5 deg$^2$. The average seeing for the central
 1 deg$^2$ field is 0$\farcs$85, while the seeing is
 $\sim$1$\farcs$00 otherwise. 
  The central field goes as deep as $g\simeq26.5$ mag and
 $u^{*}\simeq26.2$ mag at 5$\sigma$ flux limit within 3$\arcsec$ 
 aperture (or $\sim$50\% completeness), while the shallower 
 $u^*$-band data goes to 25.4 mag. 
  Our data are deep enough to detect optical counterparts of
 many IRAC and MIPS sources in the FLS region.
  We recalibrated the astrometric solution, and the final absolute
 astrometric uncertainty is at the level of 0$\farcs$1 when
 compared with SDSS point sources. With other surveys on FLS,
 the rms error is under 0$\farcs$3, and it reflects merely the
 astrometric accuracy of the other datasets. The high 
 astrometirc accuracy allows this data 
 to  be used efficiently for multislit or fiber spectroscopic
 observation of faint galaxies in the FLS.
  Using dual mode photometry, with $g$-band as the reference
 image, about 200,000 extragalactic objects are detected in
 our central 1 deg$^2$ (effective observed area reaches
 $\simeq0.94$ deg$^2$) down to $g\simeq26.5$ mag. 
  There are $\sim$50,000 sources per deg$^2$ in the shallower
 $u^*$-band images down to  $u^{*}\simeq25.4$ mag.
  Our galaxy number counts are consistent with other studies,
 and the number of spurious detections is less than 0.7\% below
 our limiting magnitudes. Our data could be used in many ways:
 for example, i) selection of high redshift objects using
 broad-band dropout technique, ii) improvement of photometric
 redshifts, and iii) investigation on rest-frame UV properties
 of high redshift infrared sources. These studies are currently
 under way.
  The mosaic images and catalogs will be available through
 a public website in near future.

\acknowledgements
  We acknowledge the FLS members for their support on this program,
 and Luc Simard and Tom Soifer for helping us to prepare the CFHT
 proposal. This work was supported by the university-institute
 cooperative research fund from the Korea Astronomy and Space
 Science Institute, and the grant No. R01-2005-000-10610-0 from
 the Basic Research Program of the Korea Science \& Engineering 
 Foundation. 
  We also acknowledge the hospitality of the Korean Institute
 of Advanced Study during the final stage of this work, and 
 the support from the School of Earth \& Environmental
 Sciences Brain Korea 21 program at Seoul National University.

{\it Facility:} \facility{CFHT (MegaCam)}

\end{document}